\documentclass{article}
\pdfoutput=1

\usepackage{arxiv}

\usepackage[utf8]{inputenc} 
\usepackage[T1]{fontenc}    
\usepackage{hyperref}       
\usepackage{url}            
\usepackage{booktabs}       
\usepackage{amsfonts}       
\usepackage{nicefrac}       
\usepackage{microtype}      
\usepackage{graphicx}
\usepackage{gensymb}
\usepackage{xcolor}
\usepackage{mathtools}
\usepackage{upgreek}
\usepackage{textcomp}
\usepackage{siunitx}

\title{On the magnetic nanostructure of a Co-Cu alloy processed by high-pressure torsion}

\author{
  Martin St\"uckler \\
  Erich Schmid Institute of Materials Science, Austrian Academy of Sciences\\
  Jahnstra{\ss}e 12, 8700 Leoben, Austria \\
  \texttt{martin.stueckler@oeaw.ac.at} 
     \And
     Christian Teichert\\
     Institute of Physics, Montanuniversit\"at Leoben\\
	 Franz-Josef-Stra\ss e 18, 8700 Leoben, Austria
	 
	 \And
	 Aleksandar Matkovi\'c\\
	 Institute of Physics, Montanuniversit\"at Leoben\\
	 Franz-Josef-Stra\ss e 18, 8700 Leoben, Austria
     \And
  Heinz Krenn \\
  Institute of Physics, University of Graz\\
  Universit\"atsplatz 5, 8010 Graz, Austria
    \And
Lukas Weissitsch\\
  Erich Schmid Institute of Materials Science, Austrian Academy of Sciences\\
  Jahnstra{\ss}e 12, 8700 Leoben, Austria 
\And
Stefan Wurster\\
  Erich Schmid Institute of Materials Science, Austrian Academy of Sciences\\
  Jahnstra{\ss}e 12, 8700 Leoben, Austria 
\And
Reinhard Pippan\\
  Erich Schmid Institute of Materials Science, Austrian Academy of Sciences\\
  Jahnstra{\ss}e 12, 8700 Leoben, Austria 
  \And
  Andrea Bachmaier\\
  Erich Schmid Institute of Materials Science, Austrian Academy of Sciences\\
  Jahnstra{\ss}e 12, 8700 Leoben, Austria 
}

\begin{document}
\maketitle
\begin{abstract}
In this study, a preparation route of Co-Cu alloys with soft magnetic properties by high-pressure torsion deformation is introduced. Nanocrystalline, supersaturated single-phase microstructures are obtained after deformation of Co-Cu alloys, which are prepared from an initial powder mixture with Co-contents above 70 wt.\%. Isochronal annealing treatments up to 400\degree C further reveal a remarkable microstructural stability. Only at 600\degree C, the supersaturated phase decomposes into two fcc-phases. The coercivity, measured by SQUID as a function of annealing temperature, remains significantly below the value for bulk-Co in all states investigated. In order to understand the measured magnetic properties in detail, a quantitative analysis of the magnetic microstructure is carried out by magnetic force microscopy and correlated to the observed changes in coercivity. Our results show that the rising coercivity can be explained by a magnetic hardening effect occurring in context with spinodal decomposition.
\end{abstract}


\section{Introduction}
Desirable magnetic properties have frequently been attributed to the nanocrystalline regime \cite{GLEITER1989223, herzer2005round}. Nanocrystalline soft magnetic materials are already commercially produced, in particular by melt spinning, yielding amorphous sheets. In additional processing steps, these sheets are stacked and exposed to annealing treatments to adjust the grain size, further tuning the magnetic properties such as saturation magnetization, coercivity, and permeability \cite{suzuki1991soft, herzer2013modern}. In contrast to this type of material synthesis, also known as bottom-up approach, the production of such nanocrystalline magnetic materials starting with coarse grained materials (top-down approach) has gained attraction recently \cite{ambrose1993magnetic, shen2005soft}. The advantage is obvious: additional processing steps, such as stacking of sheets can be omitted. Furthermore, (expensive) rare-earth elements, necessary for metallic glass formation, can be neglected. For the top-down approach, high-pressure torsion (HPT) is an attractive technique, as the prepared samples are already present in bulk form. Moreover, with this technique the microstructure can be tuned while the sample retains its shape during preparation. Being a technique of severe plastic deformation (SPD), HPT exhibits the advantage that the applied shear deformation can act continuously, i.e., any desireable amount of deformation can be applied \cite{valiev2000bulk}. This deformation can cause grain refinement with the possibility of attaining the regime of nanocrystallinity \cite{pippan2006limits}. Metastable phases can form during HPT-deformation, and may be retained even after pressure release \cite{KILMAMETOV2018337}.
Starting with conventional powders, any chemical composition can be investigated by HPT. In particular, Co-based materials have raised interest due to their low magnetostriction, favoring soft magnetic properties \cite{GLEITER1989223, o1989opportunities}. Studies dealing with the immiscible Co-Cu system have demonstrated that single phase supersaturated solid solutions can be prepared by HPT \cite{kormout2017deformation}, whereas higher Co-contents yield better soft magnetic properties \cite{stuckler2019magnetic}. As recent studies reported on failing synthesis by HPT for Co-contents above 67~wt.\% \cite{stuckler2019magnetic}, the scope of the current study is to establish a promising sample preparation route for high Co-contents yielding a single phase supersaturated solid solution. The magnetic properties of supersaturated solid solutions are highly sensitive to temperature \cite{BACHMAIER2017744}. Therefore, in this study an in-depth characterization of the temperature stability and its influence on the magnetic properties is carried out.
\section{Experimental}
Conventional powders (Co: Alfa Aesar, -22 mesh, Puratronic$^\text{\textregistered}$, 99.998\%; Cu: Alfa Aesar, -170+400 mesh, 99.9\%) were mixed to a desired composition and hydrostatically consolidated in Ar-atmosphere. The resulting coin-shaped specimen (diameter: 8~mm; thickness: 1~mm) was exposed to SPD by HPT at temperatures between ambient conditions and 500\degree C (facilitated by inductive heating of the anvils) \cite{pippan2006limits}. During HPT, a pressure of 5~GPa was applied, while using a rotational speed of 1.28~min$^{-1}$. To ensure a microstructural steady state, 50-150 numbers of turns were chosen, corresponding to a shear strain of about 1000-3000 at a radius of 3~mm. Subsequently, isochronal annealing treatments were performed in a conventional furnace for 1~h each, followed by a quick cooling in air. Vickers microhardness was measured with a used load of 500~g (HV0.5; Buehler Micromet 5100). The microstructure was investigated using scanning electron microscopy (SEM; Zeiss LEO1525). An attached energy dispersive X-ray spectrometer (EDS; Bruker e$^-$-Flash) was used to measure the samples' chemical compositions, which are given in weight percent (wt.\%) herein. The crystallographic states were characterized using X-ray diffraction (Bruker D2 Phaser) using Co-K$_\alpha$ radiation. DC-hysteresis measurements were performed with a SQUID-magnetometer (Quantum Design MPMS-XL 7) at 300~K and 8~K. Magnetic force microscopy (MFM) measurements were carried out using a Horiba France SmartSPM in two-pass mode at lift-heights between 10~nm and 20~nm using hard magnetically coated tips (coercivity 15~Oe) exhibiting tip radii of $\leq$15~nm (NANOSENSORS$^\text{{TM}}$ SSS-MFMR). For MFM measurements, sample and tip were put on the same potential, eliminating electrostatic interaction. Topographic as well as MFM scans are visualized and processed with the software Gwyddion 2.53. Topographic scans are corrected by  mean plane subtraction as well as by aligning rows. In case of MFM scans, data are analyzed as measured.
\section{Results and Discussion}
\subsection{Sample synthesis with high Co-content}
\begin{figure}[t]
\begin{center}
\includegraphics[width=0.3\linewidth]{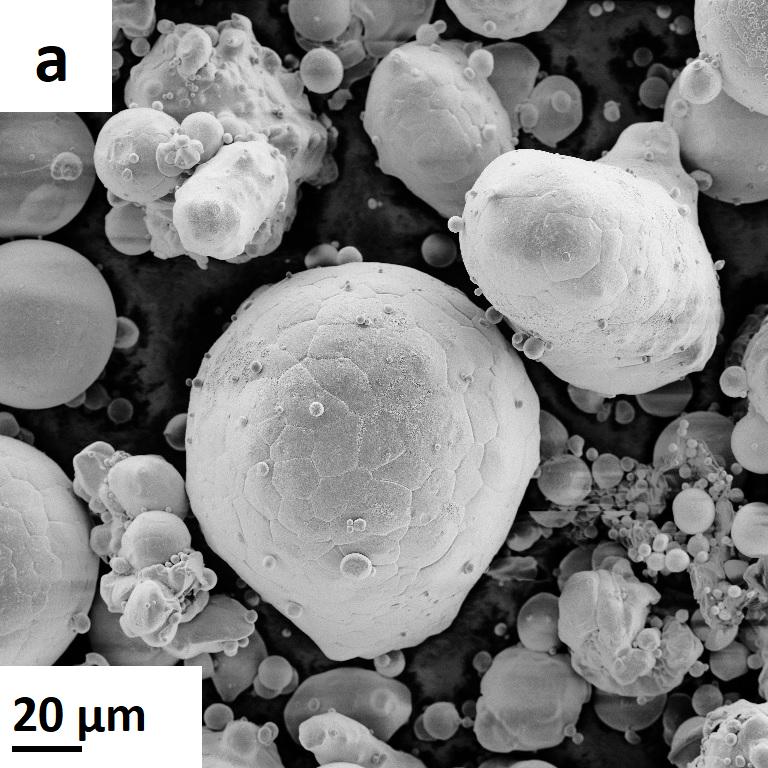}
\includegraphics[width=0.3\linewidth]{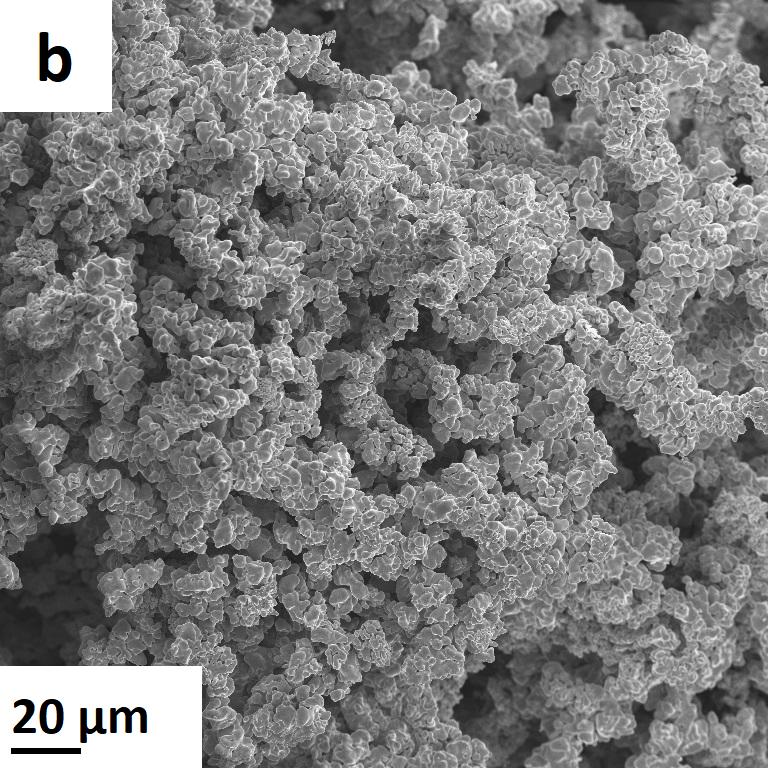}
\includegraphics[width=0.3\linewidth]{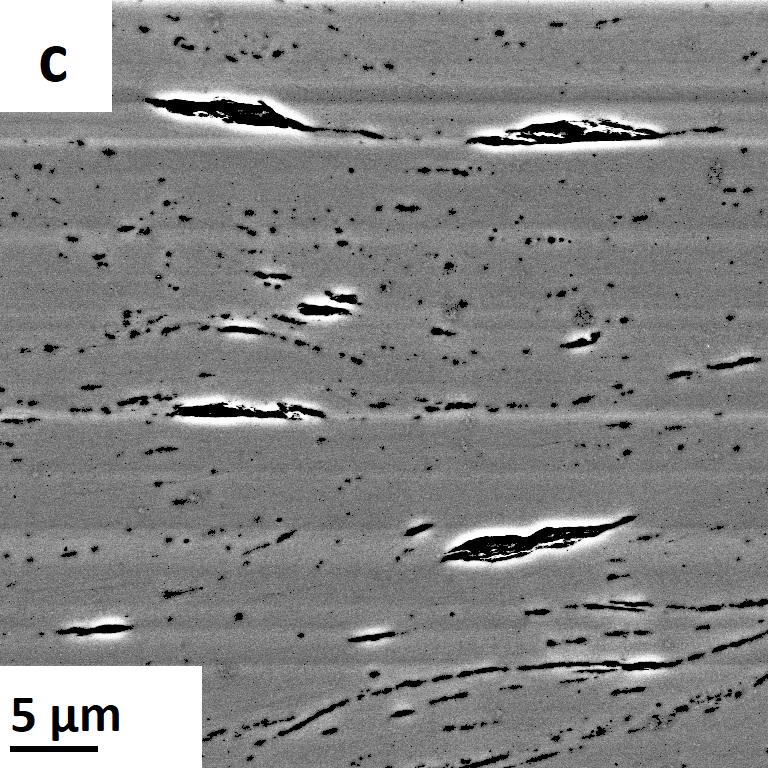}
\includegraphics[width=0.3\linewidth]{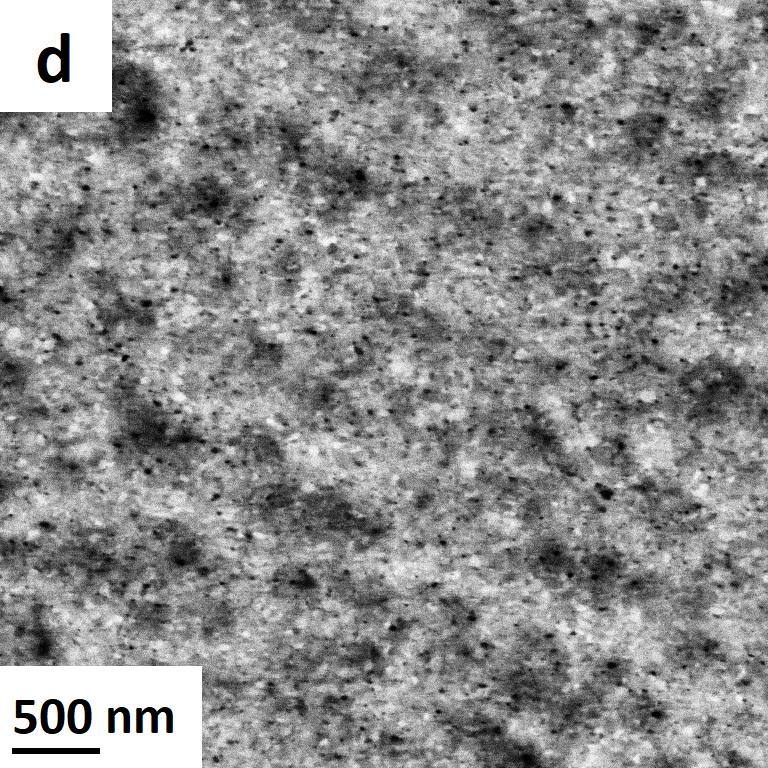}
\includegraphics[width=0.3\linewidth]{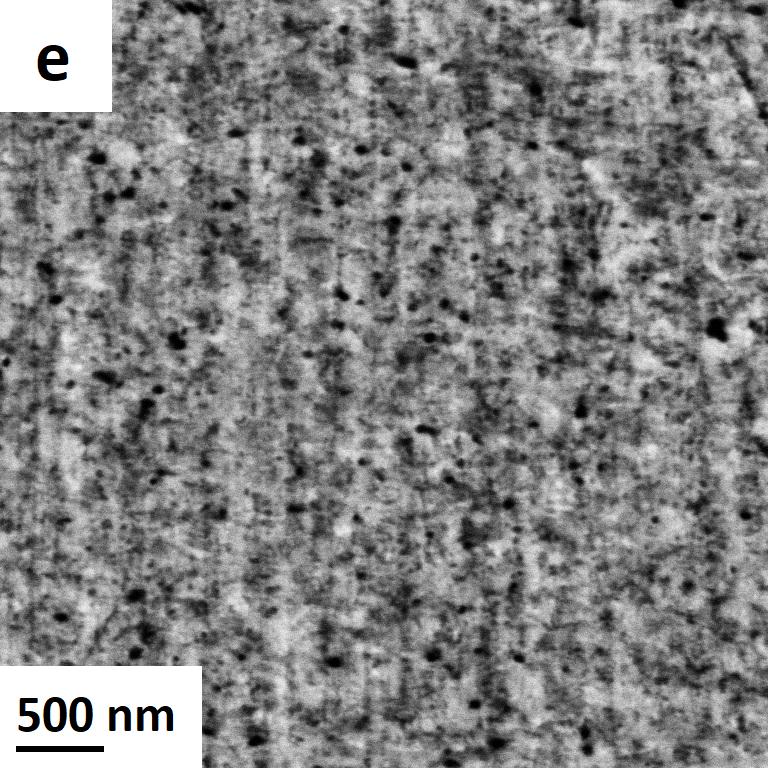}
\includegraphics[width=0.3\linewidth]{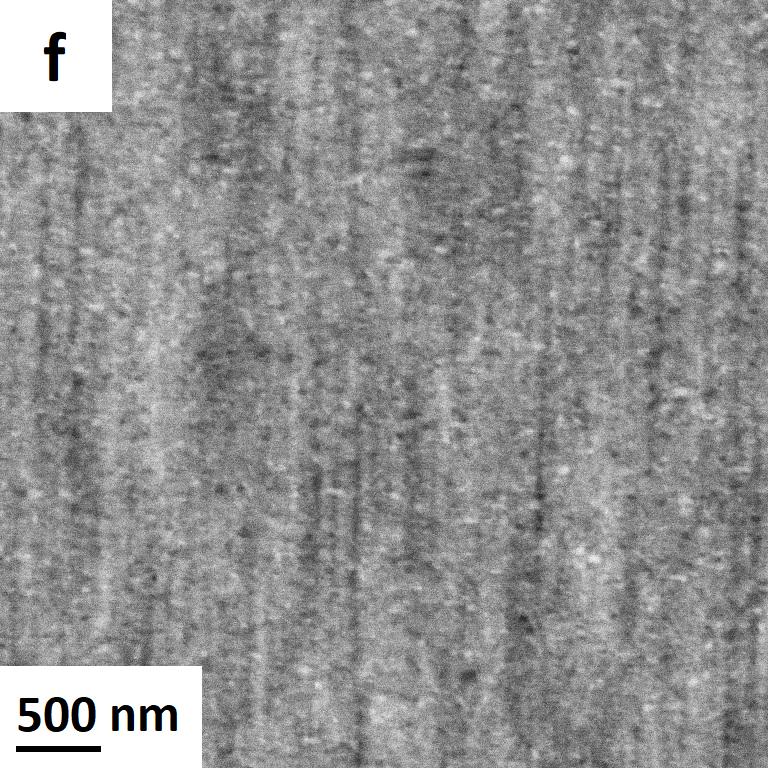}
\end{center}
\caption{SEM micrographs: (a) and (b) show the initial Cu and Co powders, respectively. (c) shows a HPT-sample deformed at RT (100 turns) consisting of Co78-Cu22 (wt.\%). (d) and (e) show Co78-Cu22 deformed at 300\degree C (100 turns) and Co93-Cu7 deformed at 500\degree C (50 turns) respectively. (f) shows a sample consisting of Co88-Cu12, which was exposed to a two-step deformation process at 500\degree C (50 turns) and RT (50 turns). Micrographs of HPT samples are taken in tangential direction at a radius of 3~mm. Please note the different scale bars.}
\label{fig:SEM1}
\end{figure}
As has recently been reported in \cite{stuckler2019magnetic}, single-phase Co-Cu samples can be prepared for intermediate Co-contents (28~wt.\%~-~67~wt.\%) by HPT at room temperature, yielding bulk samples with grain sizes in the nanocrystalline regime (77~nm-100~nm \cite{stuckler2019magnetic}). Using the same processing parameters, sample preparation with higher Co-contents ($ > $~67~wt.\%) fails due to crack formation. To obtain the desired single phase nanocrystalline microstructure at higher Co-contents, the sample synthesis is improved as presented in the following method.\\
The initial powders used in this study are shown in Fig.~\ref{fig:SEM1}(a),(b). The morphology of the Cu-powder (Fig.~\ref{fig:SEM1}(a)) appears globular, whereas the Co-powder (Fig.~\ref{fig:SEM1}(b)) shows many tiny particles agglomerating to structures with large surfaces. Room temperature deformation of Cu mixed with Co-powder shows abrasion of the harder (dark) Co particles  (Fig.~\ref{fig:SEM1}(c)) which is considered as an intermediate stage in the formation of supersaturated solid solutions \cite{bachmaier2016process}, but the formation of cracks impedes further deformation.
To overcome the crack formation limiting the development of a homogeneous microstructure, the idea is to generate the desired microstructure by two consecutive steps of HPT deformation: in the first step, HPT deformation is performed at elevated temperatures, avoiding crack formation, and generating an ultra-fine grained structure at enhanced homogeneity. In the second step, the same sample is exposed to HPT-deformation at room temperature, yielding the desired, nanocrystalline microstructure.\\
 Fig.~\ref{fig:SEM1}(d) shows an SEM micrograph of Co78-Cu22 deformed at 300\degree C for 100 turns yielding a homogeneous deformation with a complete absence of cracks. Homogeneous deformation is also reached for even higher Co-concentrations: Fig.~\ref{fig:SEM1}(e) shows Co93-Cu7 deformed at 500\degree C (50 turns). Although the deformation at elevated temperatures yields homogenization of strain distribution darker and brighter regions in the sub-\textmu m  regime are observed in the micrographs fig.~\ref{fig:SEM1}(d) and (e). Therefore, a homogeneous supersaturation is not yet reached. As the scope of this study is to reach nanocrystalline grain sizes while maintaining a high chemical homogeneity, a subsequent HPT-deformation step at lower temperatures (e.g., room temperature) is expected to yield the desired microstructure. Fig.~\ref{fig:SEM1}(f) shows a micrograph of a sample consisting of Co88-Cu12 deformed at 500\degree C (50 turns), which was subsequently exposed to an additional deformation step at room temperature (50 turns). The resulting microstructure exhibits enhanced chemical homogeneity and furthermore a smaller grain size, in comparison to Fig.~\ref{fig:SEM1}(e). As has been reported in the literature, the steady state grain size is a function of temperature \cite{pippan2010saturation}, which has been recently attributed to the deformation temperature dependent mobility of triple junctions \cite{renk2019saturation}.

\subsection{Investigations on the thermal stability}
Nanocrystalline materials are often prone to grain growth at low homologous temperatures \cite{GLEITER1989223}, causing the loss of the initial superior magnetic properties \cite{suzuki1991soft, herzer2013modern}. The thermal stability can be reduced further by a positive heat of mixing, as it is the case for the presented Co-Cu samples \cite{kormout2017deformation}. Therefore, a detailed investigation of the thermal stability of a Co72-Cu28 (wt.\%) alloy is carried out in the following, serving as an example of the Co-Cu alloys described above. The presented sample was prepared by a two-step HPT deformation process, where the first step is performed at 300\degree C (100 turns) and the second step is performed at room temperature (50 turns).
\begin{figure}[t]
\begin{center}
\includegraphics[width=0.3\linewidth]{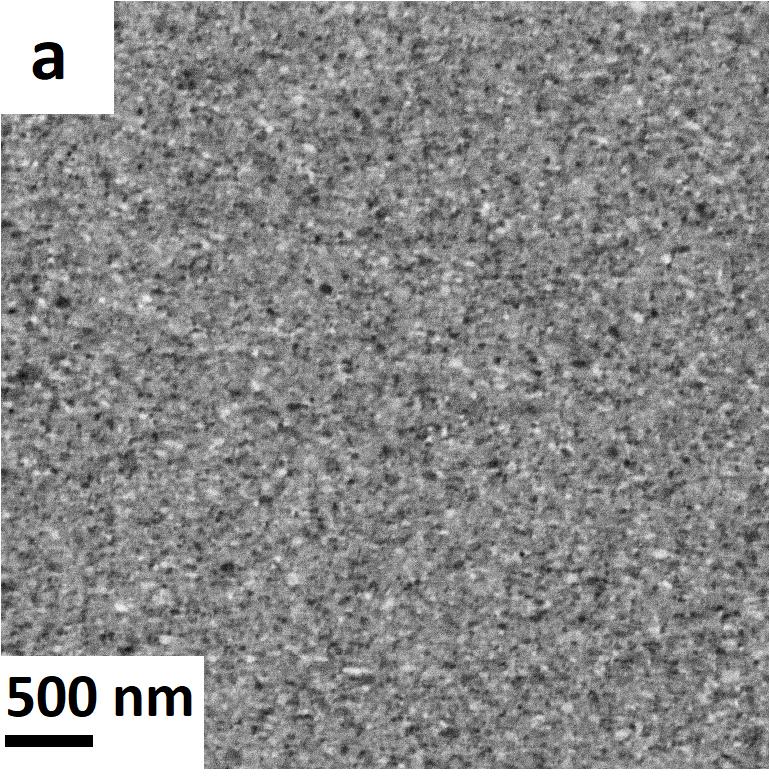}
\includegraphics[width=0.3\linewidth]{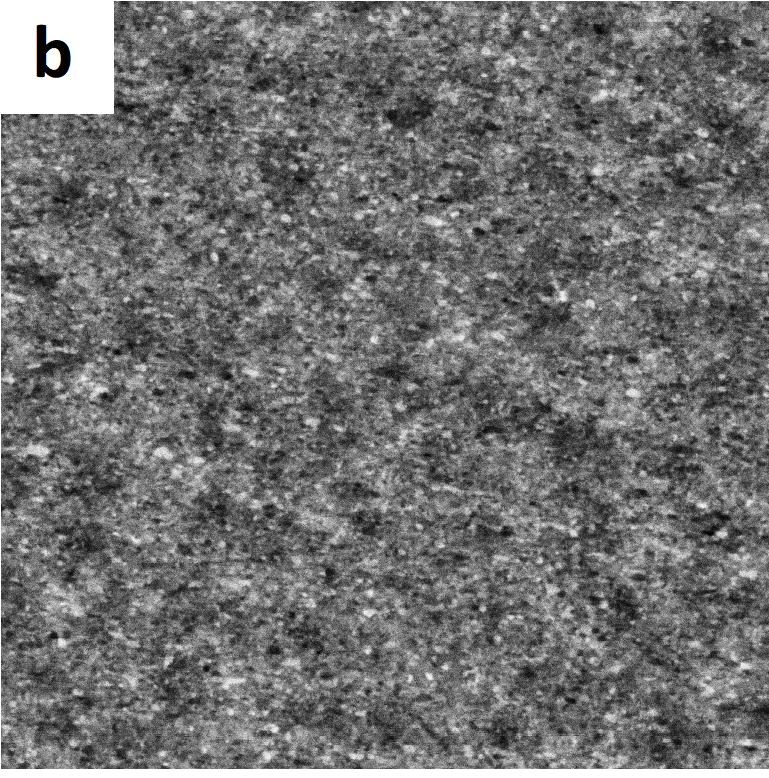}
\includegraphics[width=0.3\linewidth]{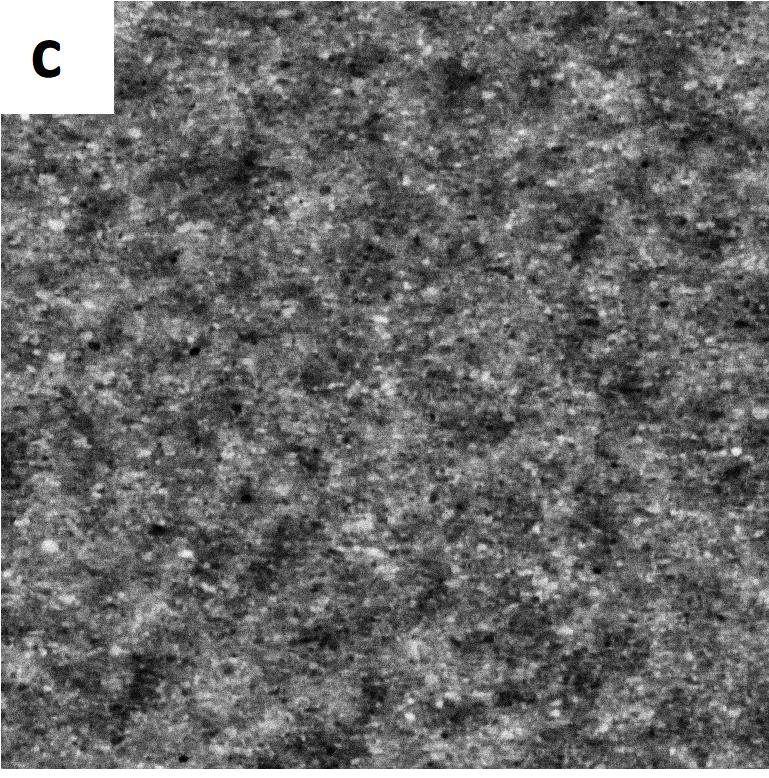}
\includegraphics[width=0.3\linewidth]{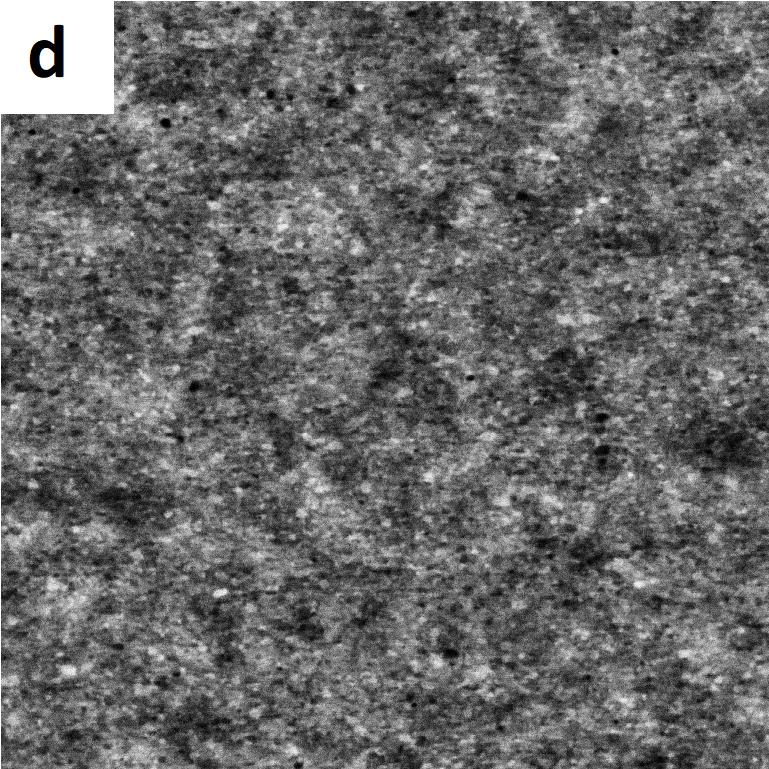}
\includegraphics[width=0.3\linewidth]{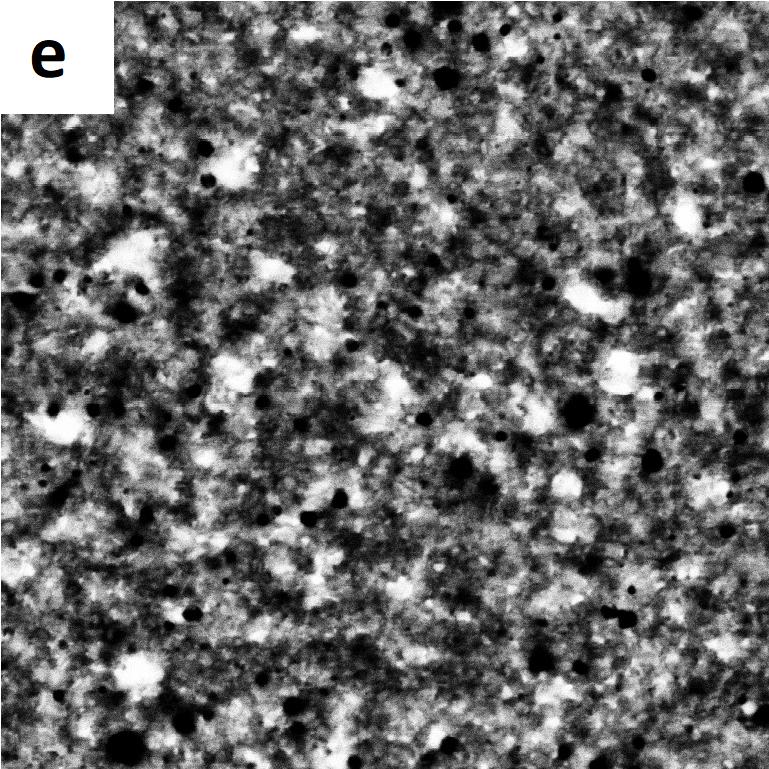}
\includegraphics[width=0.3\linewidth]{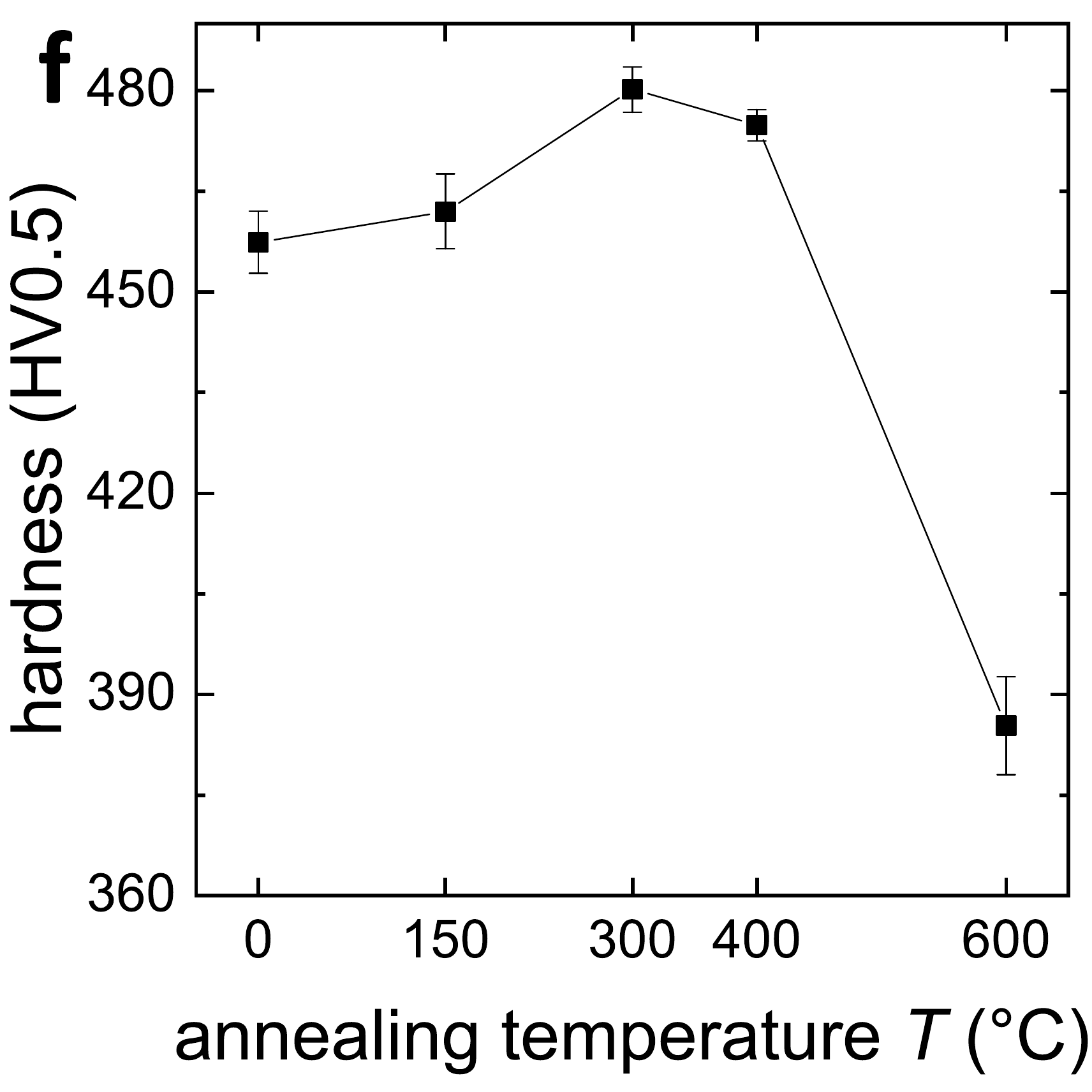}
\end{center}
\caption{SEM micrographs of (a) Co72-Cu28 (wt.\%) processed by a two-step deformation process at 300\degree C (100 turns) and subsequently at RT (50 turns). The sample in (a) has been exposed to subsequent annealing treatments  at 150\degree C (b), 300\degree C (c), 400\degree C (d) and 600\degree C (e). The scale bar in (a) applies to all micrographs. Images are taken in BSE-mode in the tangential direction. (f) shows the hardness as a function of annealing temperature.}
\label{fig:SEM2}
\end{figure}
\subsubsection{Microstructural characterization}
Figure~\ref{fig:SEM2}(a) shows the micrograph of a sample consisting of Co72-Cu28 in the as-deformed state. In comparison to the sample, which has been exposed only to deformation at 300\degree C (c.f. Fig.~\ref{fig:SEM1}(d)), the microstructure reveals an enhanced chemical homogeneity, as well as reduced grain size. The micrographs in Fig.~\ref{fig:SEM2}(b)-(e) show the sample annealed at temperatures of 150\degree C, 300\degree C, 400\degree C and 600\degree, for 1~h each. From these images, no grain growth can be observed, but chemical contrast forms slightly, indicating separation of the chemical phases for annealing treatments up to 400\degree C (Fig.~\ref{fig:SEM2}(b)-(d)). A considerable phase contrast as well as significant larger grain sizes can be observed for an annealing treatment at 600\degree C (Fig.~\ref{fig:SEM2}(e)).
Mechanical properties are very sensitive to the microstructure. Therefore, Vickers-hardness is measured as a function of annealing temperature. The mean values of several measurements between radii ,$r$, of 2~mm and 3.5~mm (according to an average of the HPT strain status) are plotted in Fig.~\ref{fig:SEM2}(f). Due to the nanocrystalline grain size as well as solid solution hardening, the hardness in the as-deformed state is enhanced with respect to the coarse grained and pure phase materials. An increasing hardness is observed for annealing treatments up to 400\degree C, which is a common behavior for nanocrystalline materials, for details see \cite{renk2015increasing}. The 600\degree C-annealed state shows the lowest hardness among all measured values, which can be explained by the largest grain size as well as by the reduction or absence of solid solution hardening.\\
\begin{figure}
\begin{center}
\includegraphics[width=0.7\linewidth]{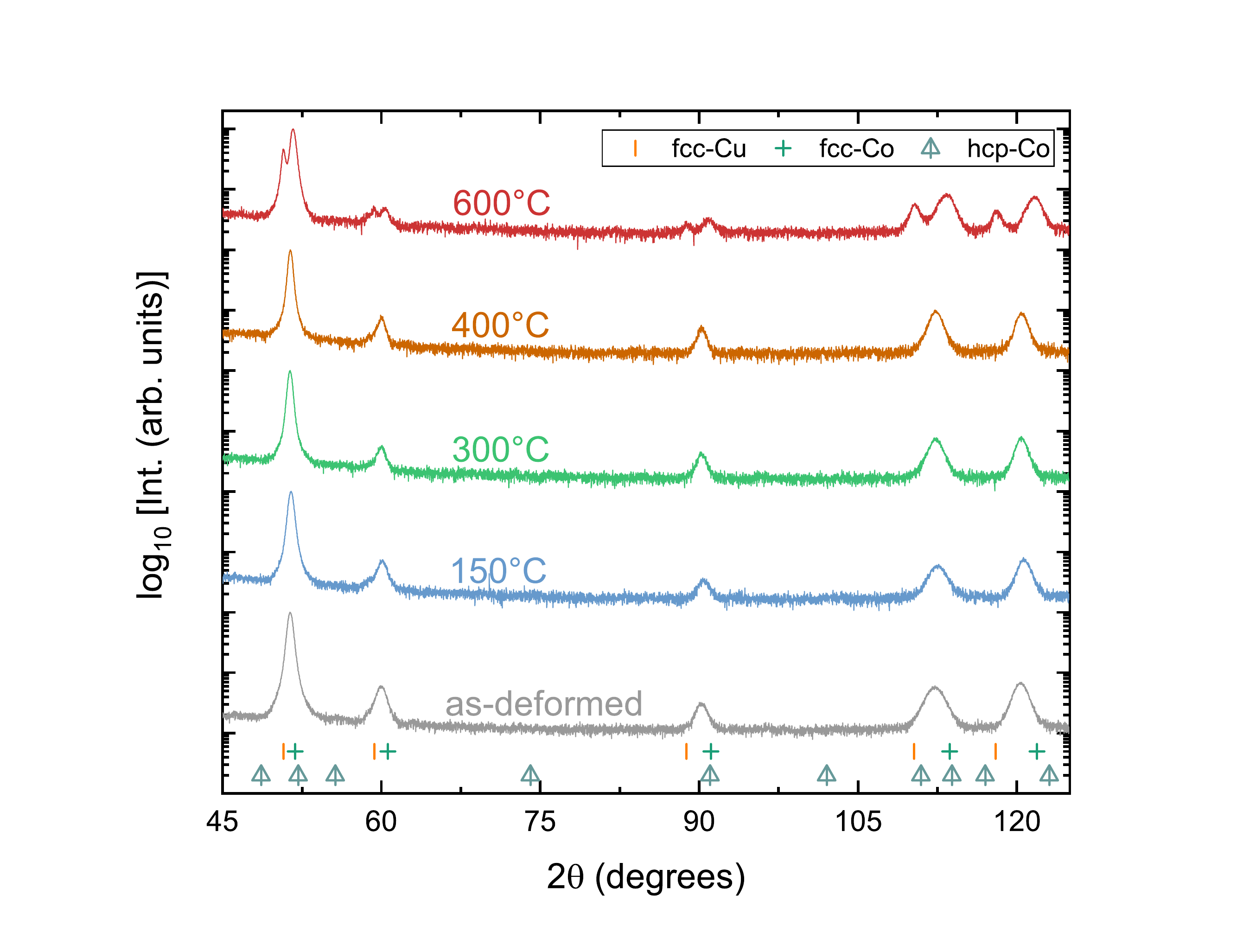}
\end{center}
\caption{XRD-pattern (Co-K$_\alpha$ radiation) of Co72-Cu28 (wt.\%) exposed to different subsequent annealing treatments (c.f. Fig.~\ref{fig:SEM2}). Peak positions of fcc-Cu, fcc-Co and hcp-Co are indicated below the spectra.}
\label{fig:xrd}
\end{figure}
XRD-measurements are presented in Fig.~\ref{fig:xrd}. In the as-deformed state, only one set of fcc-peaks is visible, showing a single-phase solid solution. Apart from slightly diminishing FWHM, no significant change in the XRD spectra can be observed for annealing treatments up to 400\degree C. A splitting into two sets of fcc-peaks is found for annealing at 600\degree C coinciding with the microstructure obtained by SEM (c.f. Fig.~\ref{fig:SEM2}(e)). From the XRD-pattern, the lattice constants are evaluated. To compensate for experimental errors, the Nelson-Riley function is used for a precise evaluation of the lattice constants \cite{nelson1945experimental}.
\begin{equation}
\frac{\Delta a}{a} \propto \frac{cos^2\theta}{sin\theta} + \frac{cos^2\theta}{\theta}
\label{eq:nelson-riley}
\end{equation}
[Eq.~(\ref{eq:nelson-riley})] is used for evaluating the lattice constant $a$, with $\theta$ being the diffraction angle of the maxima. The y-intercept of a linear regression represents $\Delta a \: = \: 0$ and leads to a precise determination in $a$. The resulting lattice constants are displayed in Fig.~\ref{fig:lattice_const}, showing slight changes for annealing treatments up to 400\degree C, arising most likely from compositional fluctuations in powder metallurgical sample preparation. At 600\degree C, two different fcc-phases are identified, with the lattice constants being very close to the pure elements. However, it can not be concluded that both phases are chemically pure, i.e, some small amounts of Co might persist in the Cu-phase and vice versa. Furthermore, the Co-contents $x_{Co}$~(at.\%) are calculated for each $a$ according to Vegard's law \cite{vegard1921konstitution}.
\begin{equation}
a=x_{Co} \cdot a_{Co} + (1-x_{Co}) \cdot a_{Cu}
\label{eq:vegard}
\end{equation}
For the above estimation of Co-contents, the lattice constant are taken as in the following: $a_{Cu}=3.615 \: \mathring{\text{A}}$, $a_{Co}=3.544 \: \mathring{\text{A}}$. In Fig.~\ref{fig:lattice_const}, the Co-contents according to Vegard's law are indicated.\\
Although the single-phase nanocrystalline microstructure was determined to be stable up to 400\degree C, it should be stated, that this value might not be achieved for different Co-Cu ratios. Actually, in several studies dealing with SPD-processed materials the temperature limit for single phase supersaturated microstructures was reported to be extremely sensitive to the particular composition investigated \cite{KILMAMETOV2018337, kriegel2019thermal}.
\begin{figure}
\begin{center}
\includegraphics[width=0.7\linewidth]{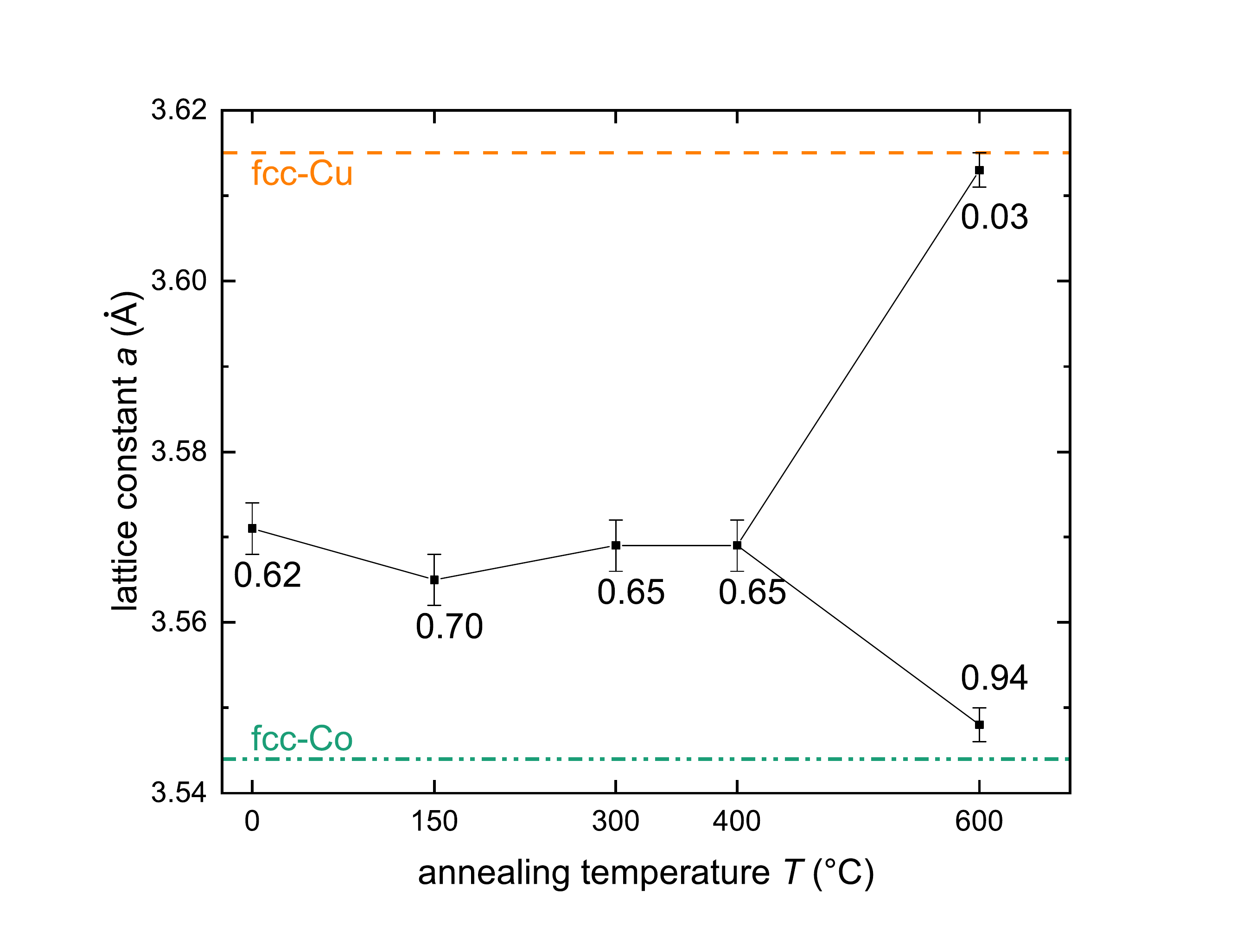}
\end{center}
\caption{Lattice constants as evaluated with the Nelson-Riley function [eq.~(\ref{eq:nelson-riley})] from XRD-patterns. The Co-content (at.\%) is calculated using Vegard's law for each lattice constant (orange dashed line: $a_{Cu}=3.615 \: \mathring{\text{A}}$, green dashed line: $a_{Co}=3.544 \: \mathring{\text{A}}$).}
\label{fig:lattice_const}
\end{figure}

\subsubsection{DC-magnetic properties}
\begin{figure}
\begin{center}
\includegraphics[width=0.7\linewidth]{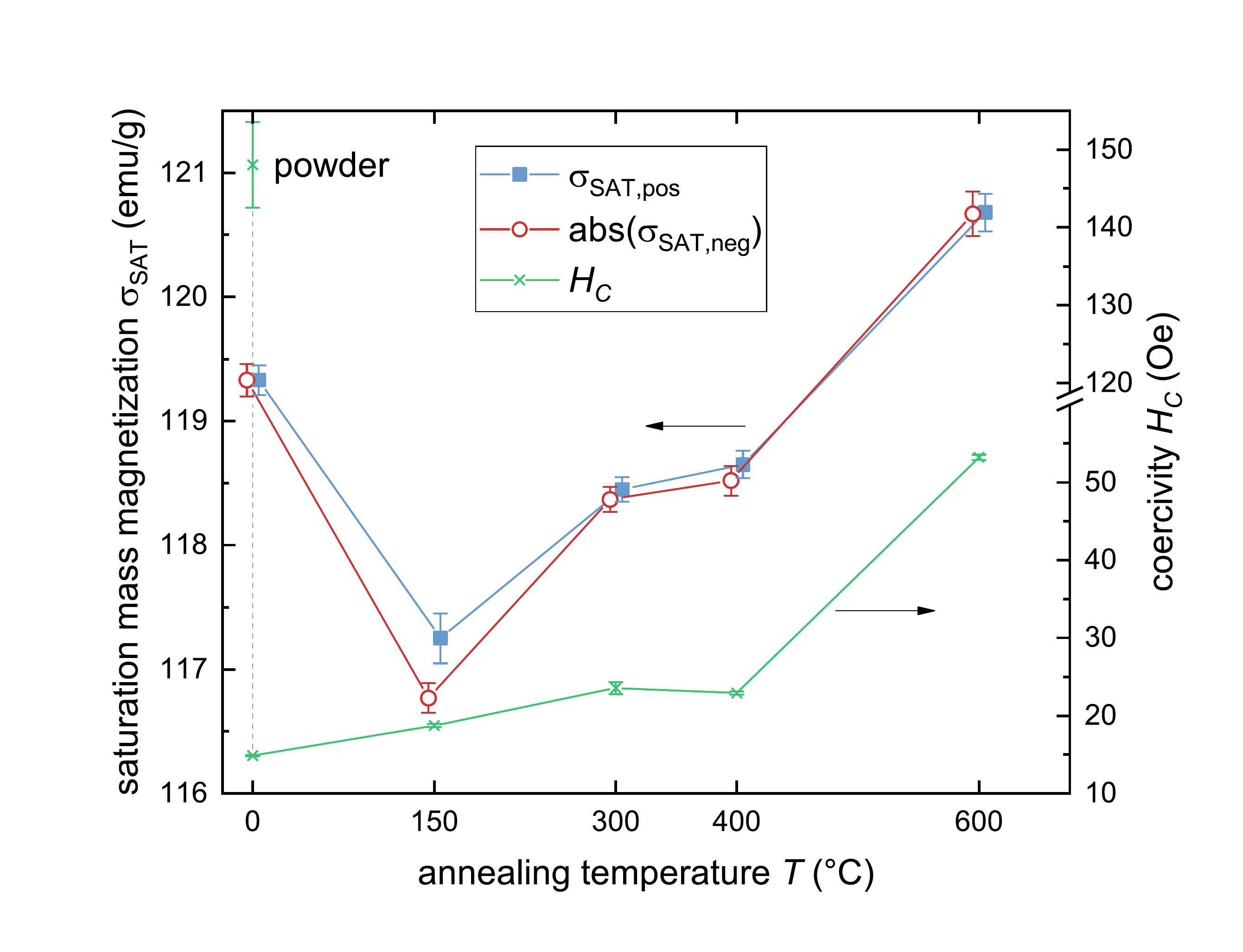}
\end{center}
\caption{Saturation mass magnetization $\sigma_{SAT}$ (at $300$~K) and coercivity $H_C$ (at $8$~K) evaluated as a function of annealing temperature. The data are slightly shifted in $x$ for better visibility. The saturation magnetization of fcc-Co is $166.1$~emu $\cdot$ g$^{-1}$. The coercivity of the initial Co-powder is also indicated. Note the break in the right y-axis.}
\label{fig:msat_Hc}
\end{figure}

In \cite{stuckler2019magnetic}, it has been proposed that an increasing Co-content yields higher saturation magnetization as well as lower coercivity. To examine this presumption, hysteresis measurements using SQUID are performed at $300$~K and $8$~K between $-7$~T and $+7$~T. DC-hysteresis curves are shown in Fig.~S.2. The approach to saturation magnetization is monitored by plotting the mass magnetization $\sigma$ versus the inverse external field $H^{-1}$ between $2$~T and $7$~T, whereas the y-intercept of a linear regression yields $\sigma_{SAT}$ at $H\: = \: \infty$, i.e., the saturation mass magnetization, which is plotted in Fig.~\ref{fig:msat_Hc}. The positive as well as the negative saturation magnetization is evaluated. Fig.~\ref{fig:msat_Hc} shows changes in $\sigma_{SAT}$ upon annealing. It is known that the magnetic moment of fcc-Co ($166.1$~emu$\cdot$g$^{-1}$) deviates from the magnetic moment of hcp-Co ($162.5$~emu$\cdot$g$^{-1}$). 
As revealed in the XRD-pattern (c.f. Fig.~\ref{fig:xrd}), Co is present in an fcc configuration in all investigated states. It is therefore more likely that the observed changes in $\sigma_{SAT}$ up to an annealing temperature of 400~\degree C arise from compositional fluctuations, which can be also seen in the XRD-pattern to some extent and are a common problem in powder metallurgy. A change in $\sigma_{SAT}$ between 400\degree C and 600\degree C can be attributed to chemical demixing, as the supersaturated Co-Cu phase exhibits a suppressed magnetic moment \cite{childress1991reentrant}. Assuming the magnetic moment for fcc-Co to be $166.1$~emu$\cdot$g$^{-1}$  an evaluation of  the mean Co-content from $\sigma_{SAT}$ yields $71 \: \pm \: 1$~wt.\%. This is in good accordance with the results from XRD-measurements, taking into account the completely different approaches for these quantification methods.\\
For an enhanced resolution in the determination of the coercivity $H_C$, hysteresis loops are measured at $8$~K. It should be stated that errors in the applied $H$-field lead to a deviation in the determination of $H_C$. But as this is a reversible artifact \cite{qdusa1500-011}, the relative evolution of $H_C$ is unaffected. $H_C$ as determined from hysteresis measurements is also plotted in Fig.~\ref{fig:msat_Hc} for all measured states. $H_C$ increases with increasing annealing temperature, in contrast to the classical assumption that large grain sizes (possessing multidomain structure) favor soft magnetic properties. The observed behavior of $H_C$ corresponds rather to random anisotropy \cite{herzer2013modern, kronmuller2003micromagnetism}. However, it should be noted that a prerequisite for random anisotropy is for the magnetic exchange length to exceed the grain size, which seems not to be the case when referring to micromagnetic constants of bulk-Co. Another important aspect is the huge change in coercivity between the initial powder and the as-deformed state. $H_C$ of the initial powder is also not reached for annealing treatments at 600\degree C, revealing the strong influence of the microstructure on the coercivity. Furthermore, a reduced magnetocrystalline anisotropy is assumed for fcc-Co in comparison to hcp-Co, which is present in the initial powder \cite{jamet2001magnetic}. Yet the mentioned points do not suffice to quantitatively describe the observed changes in $H_C$. A similar temperature behavior of the coercivity has also been reported for SPD-processed Co26-Cu74 (at.\%) \cite{BACHMAIER2017744}, but in contrast, in the present study a significantly smaller coercivity is measured and no indications of dilute ferromagnetic phase nor superparamagnetism were found. This is most likely due to the enhanced Co-content in our study forming percolating ferromagnetic structures rather than isolated clusters.

\subsubsection{Magnetic force microscopy}
\begin{figure}
\begin{center}
\begin{tabular}{cc}
  \includegraphics[width=0.3\linewidth]{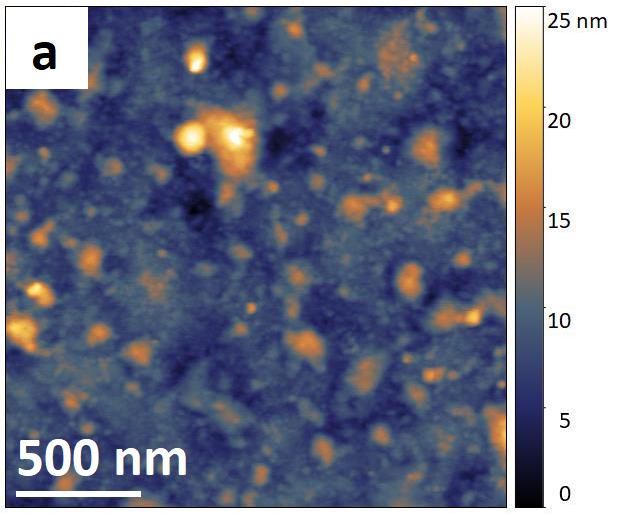} &   \includegraphics[width=0.3\linewidth]{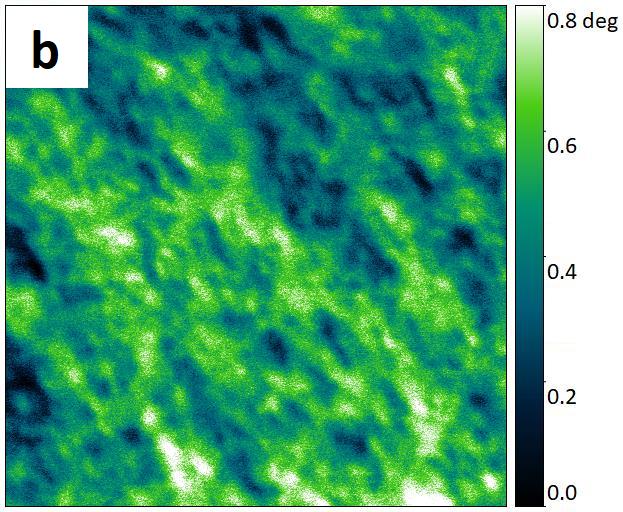} \\
  \includegraphics[width=0.3\linewidth]{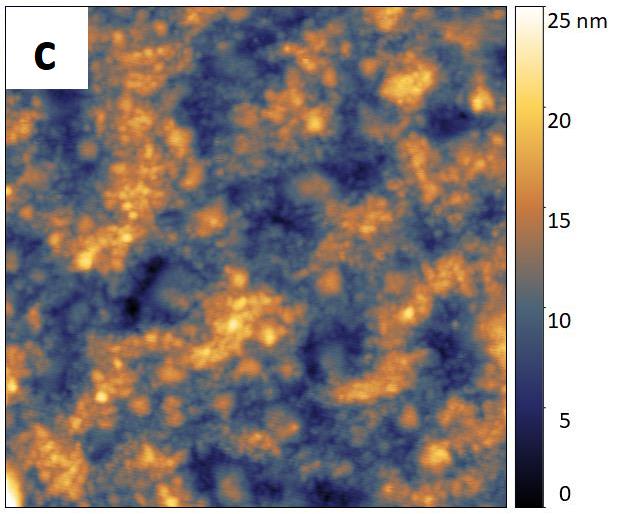} &   \includegraphics[width=0.3\linewidth]{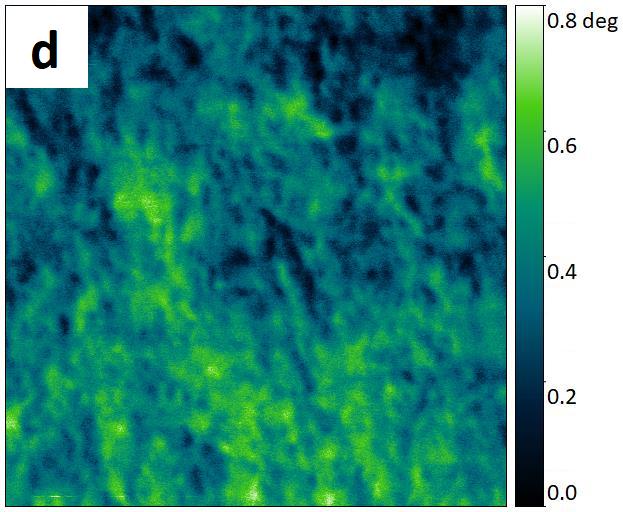} \\
\end{tabular}
\end{center}
\caption{2\textmu m x 2\textmu m AFM scans of (a) as-deformed state and (c) 300\degree C annealed state. The corresponding MFM scans of the as-deformed and 300\degree C annealed state are shown in (b) and (d) respectively. The axial direction of the HPT specimen points out of the plane, the shear direction is in horizontal direction. The lateral scale bar in (a) applies to all scans. The minimum height and phase signal values are shifted to zero for visualization purposes.}
\label{fig:MFM}
\end{figure}

For an in-depth  understanding of the observed DC-magnetic properties, a quantitative analysis of the magnetic microstructure is carried out. MFM measurements are performed in two-pass intermittant-contact (tapping) mode, i.e., in the first pass the topography is measured, whereas in the second pass the magnetic stray field is measured by keeping the tip-to-sample distance constant at an increased lift-height to minimize short-range interactions contribution in the phase lag of the cantilever oscillations in the second pass, i.e., to measure mainly magnetic force contribution to the phase lag of the cantilever in the second pass. The magnetic microstructure is measured in axial sample direction. Fig. ~\ref{fig:MFM} shows representative topography and MFM measurements of the as-deformed and the 300\degree C annealed state. The topography (Fig.~\ref{fig:MFM}(a),(c)) appears very smooth in both states and does not show any correlation to its respective MFM scan (Fig.~\ref{fig:MFM}(b),(d)). An irregular appearance of the domains is visible in both MFM measurements, most likely arising from sample synthesis by SPD causing residual stresses \cite{andreeva2016study}. The domain sizes are in the range of 100~nm corresponding to the grain size in the as-deformed state \cite{stuckler2019magnetic}, indicating single-domain behavior.
To qualitatively characterize the morphology and magnetic domain sizes for each annealing temperature, we employ the following procedure which has been introduced for a comprehensive surface roughness characterization \cite{zhao2000characterization, teichert2002self}  but can also be applied to the MFM signal \cite{teichert2009ion}. This procedure considers so-called auto-correlation and height-height correlation functions \cite{zhao2000characterization}. For isotropic samples as in our case it is sufficient to perform a one-dimensional \mbox{analysis}. The auto-correlation function $C(x)$ is calculated for each scan line of the image \cite{zhao2000characterization}. 
\begin{equation}
\label{eq:C_x}
\centering
C(x)=<[z(x_0+x) \: - <z>][z(x_0) \: -  <z>]>
\end{equation} 
where $z$ stands either for the height in nm or the phase signal of the MFM image in degrees. For a self-affine surface with a cut-off \cite{zhao2000characterization, teichert2002self} $C(x)$ can be described by
\begin{equation}
\label{eq:C_x_fit}
\centering
C(x)=\sigma^2 \cdot exp [-(|x|/\xi)^{2\alpha}]
\end{equation}
with $\sigma$ being the root mean square roughness (for MFM $\sigma$ represents the standard deviation of the average magnetic signal which is related to the out-of-plane magnetization contrast). $\xi$ represents the lateral correlation length, i.e., the length scale within the height or the MFM signal of two points correlate. The Hurst parameter $\alpha$ represents the ``jaggedness'' of the surface or the magnetic signal and ranges from 0 to 1, where small values in $\alpha$ represent sudden fluctuations in height. Another function describing surface statistics is the height-height correlation function $H(x)$ \cite{zhao2000characterization}.
\begin{equation}
\label{eq:H_x}
\centering
H(x)=<[z(x_0+x)-z(x_0)]^2>
\end{equation}
with the asymptotic behavior
\begin{equation}
\label{eq:H_x_fit}
H(x) = x^{2\alpha} \: \text{for} \: x \ll \xi
\end{equation}
\begin{figure}
\begin{center}
\begin{tabular}{cc}
\includegraphics[width=0.3\linewidth]{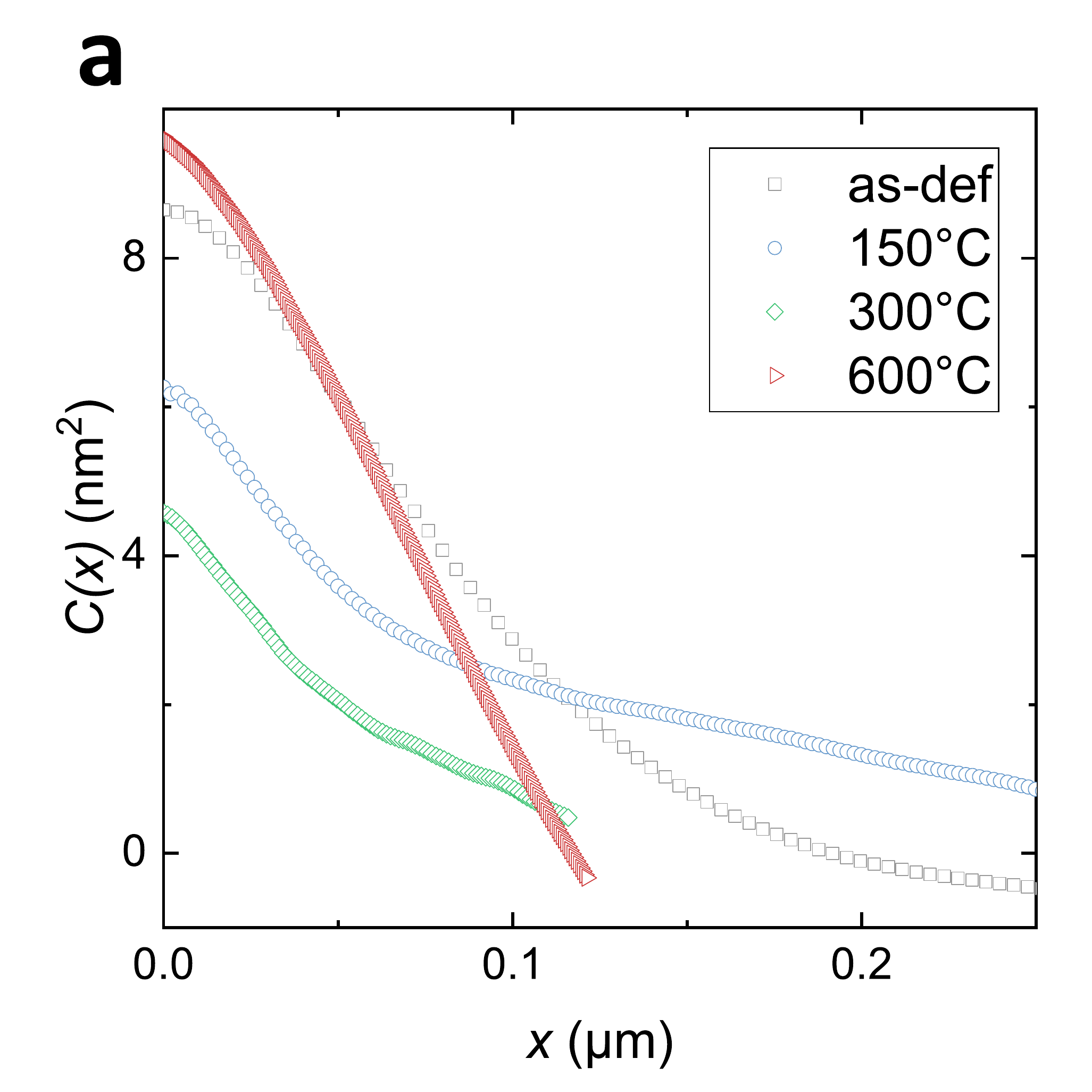} & \includegraphics[width=0.3\linewidth]{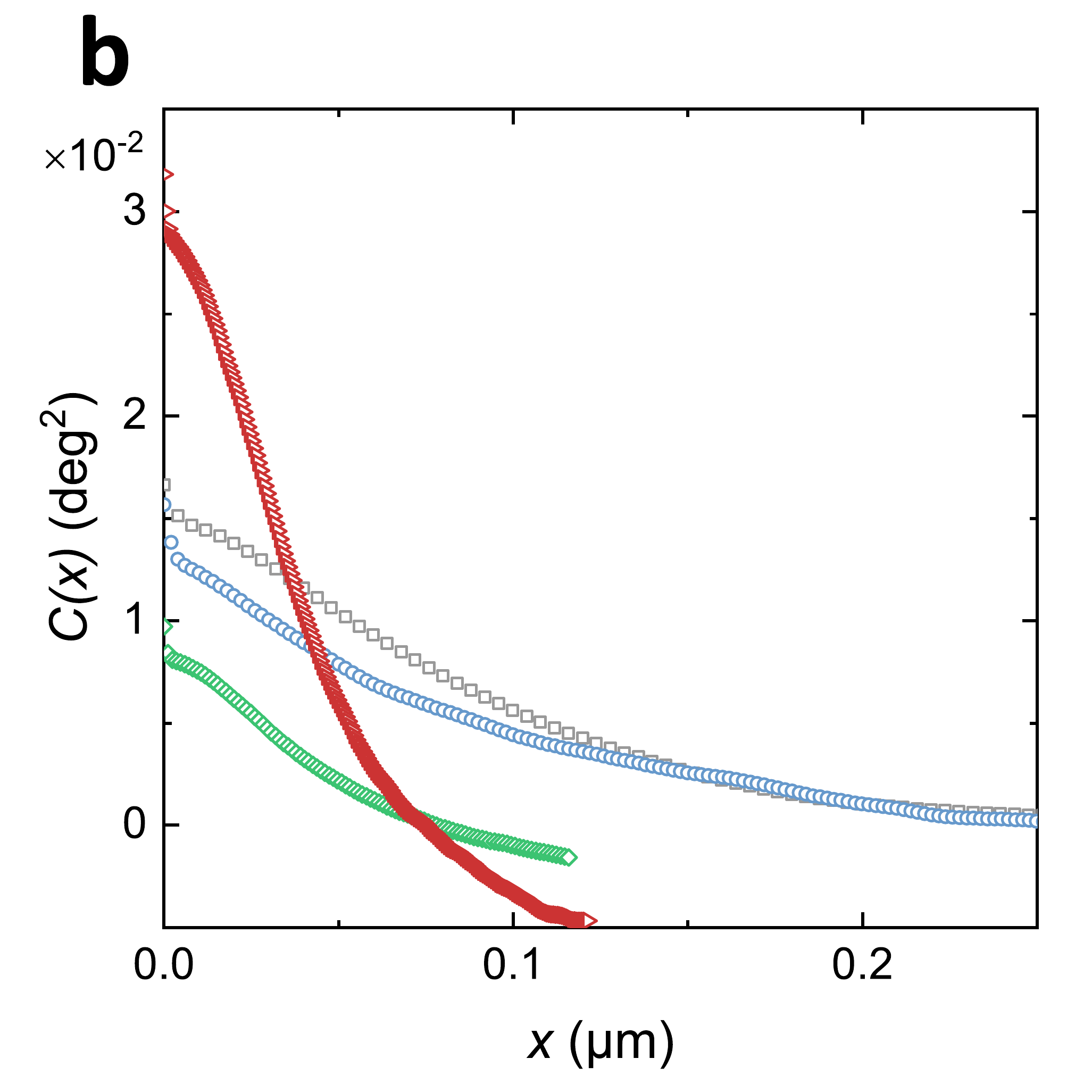} \\
\includegraphics[width=0.3\linewidth]{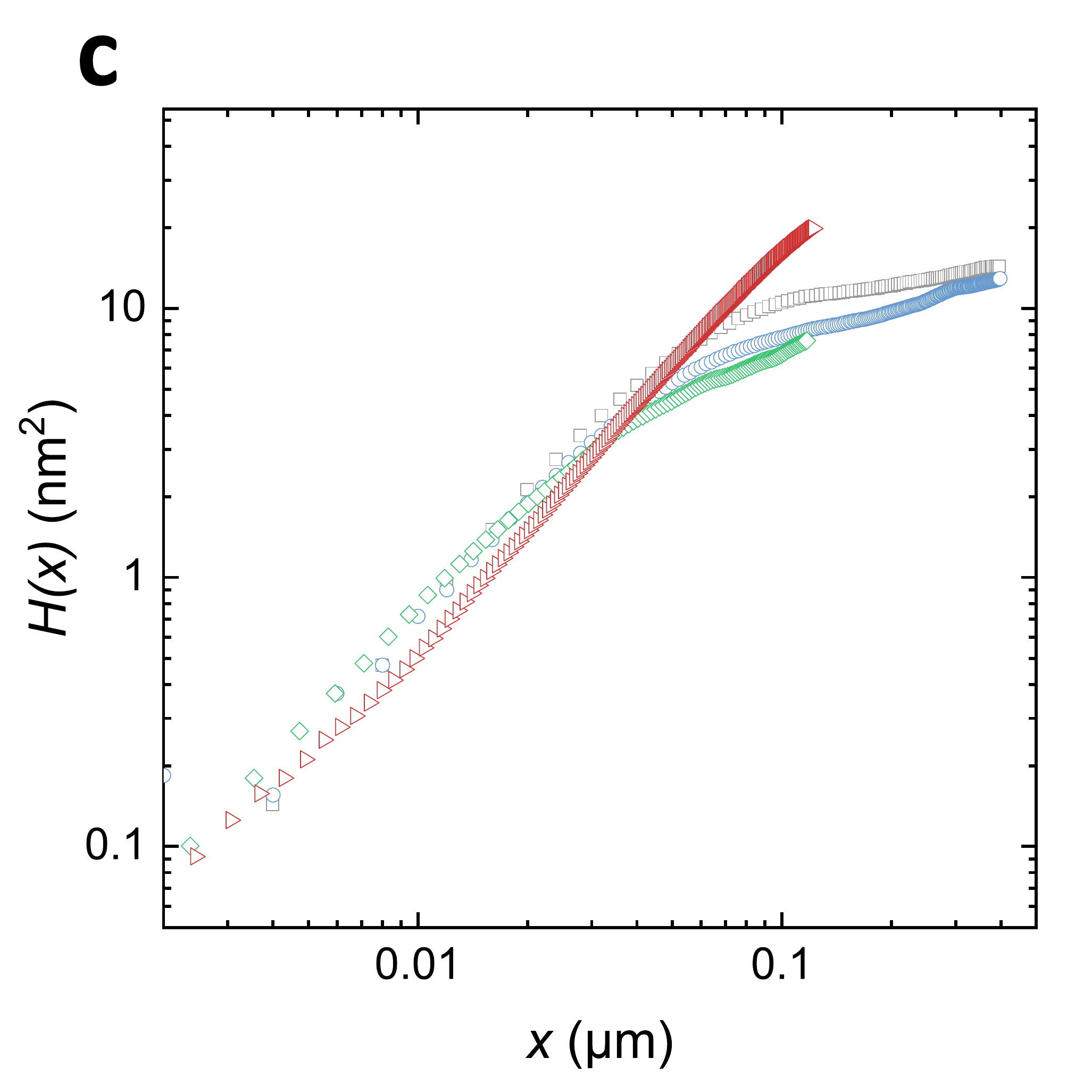} & \includegraphics[width=0.3\linewidth]{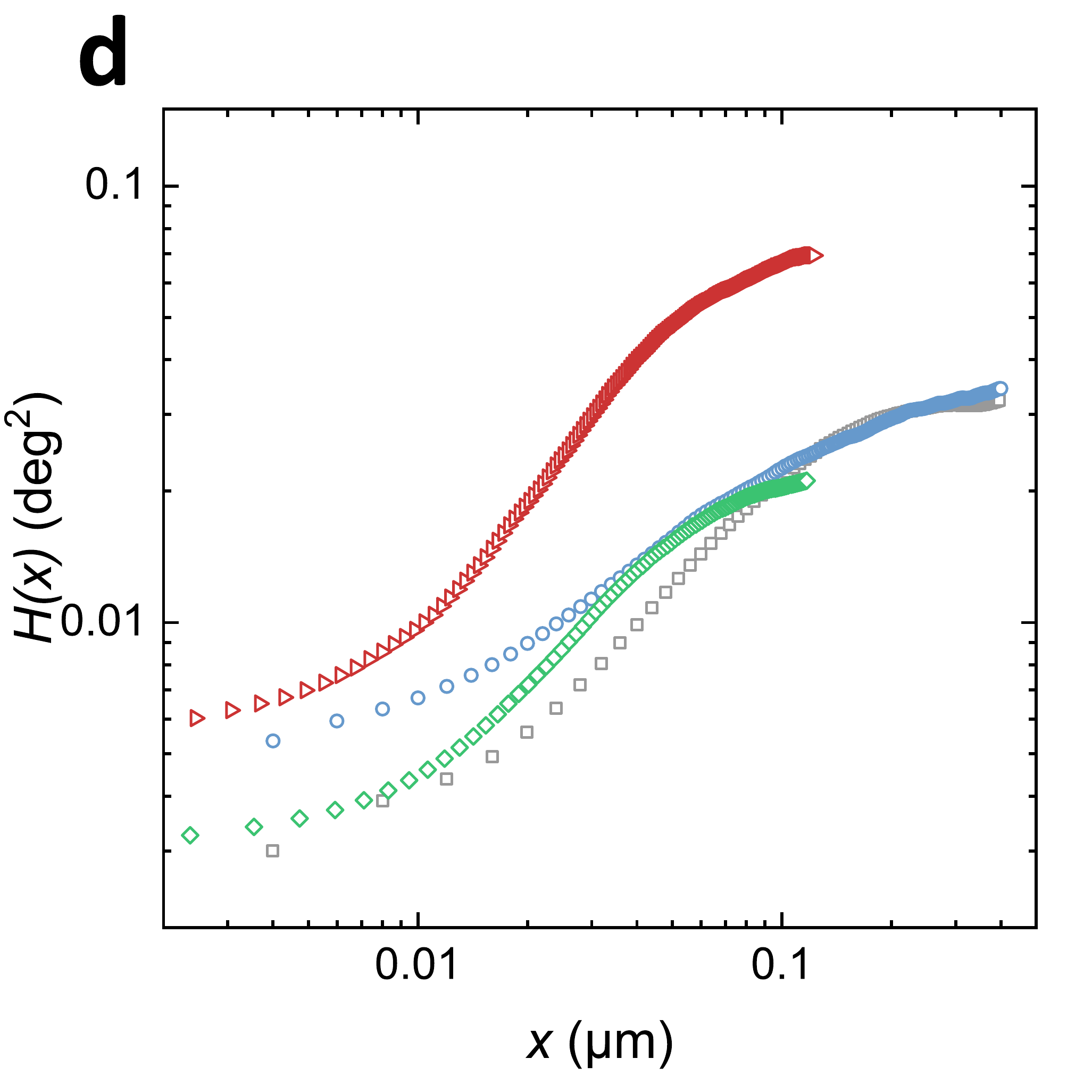}
\end{tabular}
\end{center}
\caption{Quantitative analysis of AFM and MFM-scans. (a) and (b) shows auto-correlation functions $C(x)$ [eq.~\ref{eq:C_x}] of topographic scans and MFM scans respectively. (c) and (d) shows height-height correlation functions $H(x)$ [eq.~\ref{eq:H_x}] of topographic scans and MFM scans respectively.}
\label{fig:MFM_quant}
\end{figure}
In the following, both the auto-correlation function and the height-height correlation function are calculated for the MFM scans \cite{teichert2009ion}. To disentangle the magnetic information from topography, these functions are also evaluated for the topographic (first-pass) scans. At least 6 individual scans are evaluated for each state to enhance statistical significance. Fig.~\ref{fig:MFM_quant} shows representative $C(x)$ and the $H(x)$ curves evaluated for the the topographic signal and MFM signal. Note that the data are plotted in the statistically significant range for small values in $x$. Different shapes can be identified in the presented $C(x)$. In contrast to the topographic $H(x)$ (Fig.~\ref{fig:MFM_quant}(c)),  two slopes can be identified in magnetic $H(x)$ (Fig.~\ref{fig:MFM_quant}(d)), representing a multi-fractal behavior \cite{teichert2009ion}. Therefore, a superposition of two self-affine fits \cite{tolan1998evidence} yields a better description of the magnetic $C(x)$:
\begin{equation}
\label{eq:C_x_fit2}
\centering
C(x)=\sigma^2 (c \cdot exp [-(|x|/\xi_1)^{2\alpha_1}] + (1-c) \cdot exp [-(|x|/\xi_2)^{2\alpha_2}])
\end{equation}
with $c$ being a proportionality constant (0~$\leq \: c \: \leq$~1). In tab.~S.1, the mean values of the fit parameters of $C(x)$ and $H(x)$ are listed. The proportionality constant $c$ weights the second term ($\xi_2$ and $\alpha_2$) in [eq.~\ref{eq:C_x_fit2}] stronger. The analysis of the magnetic microstructure focuses therefore on the second term in [eq.~\ref{eq:C_x_fit2}]. Fig.~\ref{fig:xi}(a) presents the magnetic correlation lengths ($\xi_1$, $\xi_2$) and the topographic correlation length ($\xi_{TOPO}$) as a function of annealing temperature. Magnetic $\xi_2$ shows a correlation with annealing temperature, whereas for the topographic lateral correlation length $\xi_{TOPO}$ such a correlation cannot be observed. We can therefore conclude that $\xi_2$ is uniquely attributed to the magnetic microstructure. In particular $\xi_2$ is (75 $\pm$ 34)~nm in the as-deformed state, corresponding to the grain size, as obtained for similar materials \cite{stuckler2019magnetic} and therefore showing single-domain behavior. $\xi_2$ decreases as a function of annealing temperature, i.e., upon annealing the domain size decreases causing a magnetic hardening.\\
In the following, the coercivity measured by SQUID-magnetometry is put into relation with $\xi_2$. 
It is important to keep in mind that the magnetic $\xi_2$ decreases with increasing annealing temperature, meaning that in Fig.~\ref{fig:xi}(b) rising annealing temperature shifts from the right to the left. When referring to annealing treatments it seems unreasonable that higher annealing temperatures cause smaller magnetic feature sizes, but as it was shown in Fig.~\ref{fig:SEM2} the microstructure does not show any change in grain size up to 400\degree C. As the as-deformed state is very far from thermodynamical equilibrium, it is more likely that the change in $\xi_2$ arises from a demixing process causing the magnetic Co-phase to agglomerate. Similar phase separation processes have already been observed in the Co-Cu system and have been attributed to spinodal decomposition \cite{bachmaier2015phase}. The same seperation process could be responsible for the improved magnetoresistive behavior of annealed Cu-Co HPT-materials \cite{wurster2020GMR}. In the 600\degree C annealed sample, a demixed state is present with the highest coercivity and the smallest magnetic $\xi_2$.\\
It is important to mention that, although the magnetic hardening can be attributed to single-domain behavior, the critical diameter for single-domain particle size is in the range of 50~nm \cite{kronmuller2003micromagnetism}. However, this value refers to bulk hcp-Co and one has to keep in mind that Co neither is present in its hcp-phase nor chemically pure, i.e., further investigations on the micromagnetic properties of such supersaturated fcc-Co phases need to be carried out.
\begin{figure}
\includegraphics[width=0.5\linewidth]{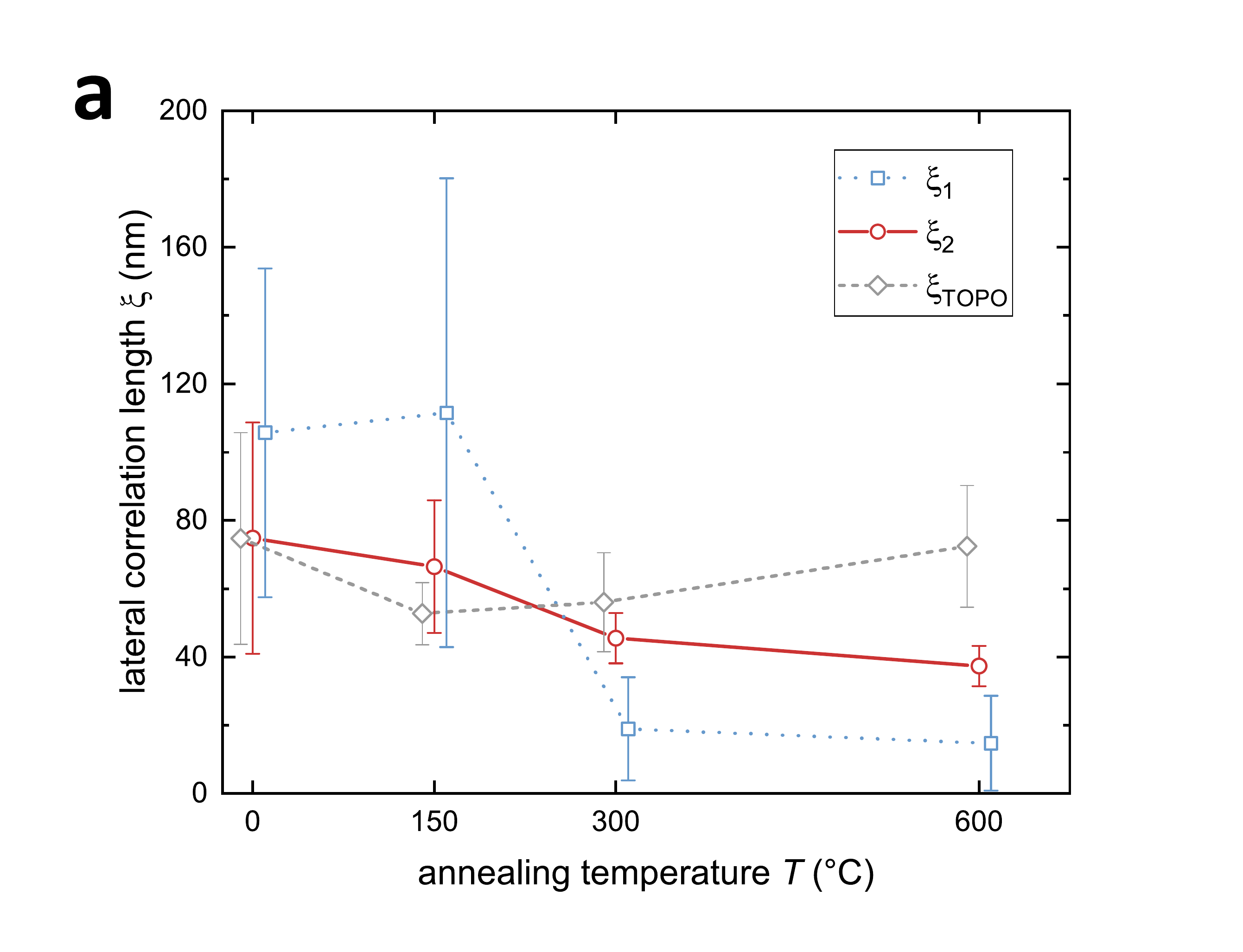}
\includegraphics[width=0.5\linewidth]{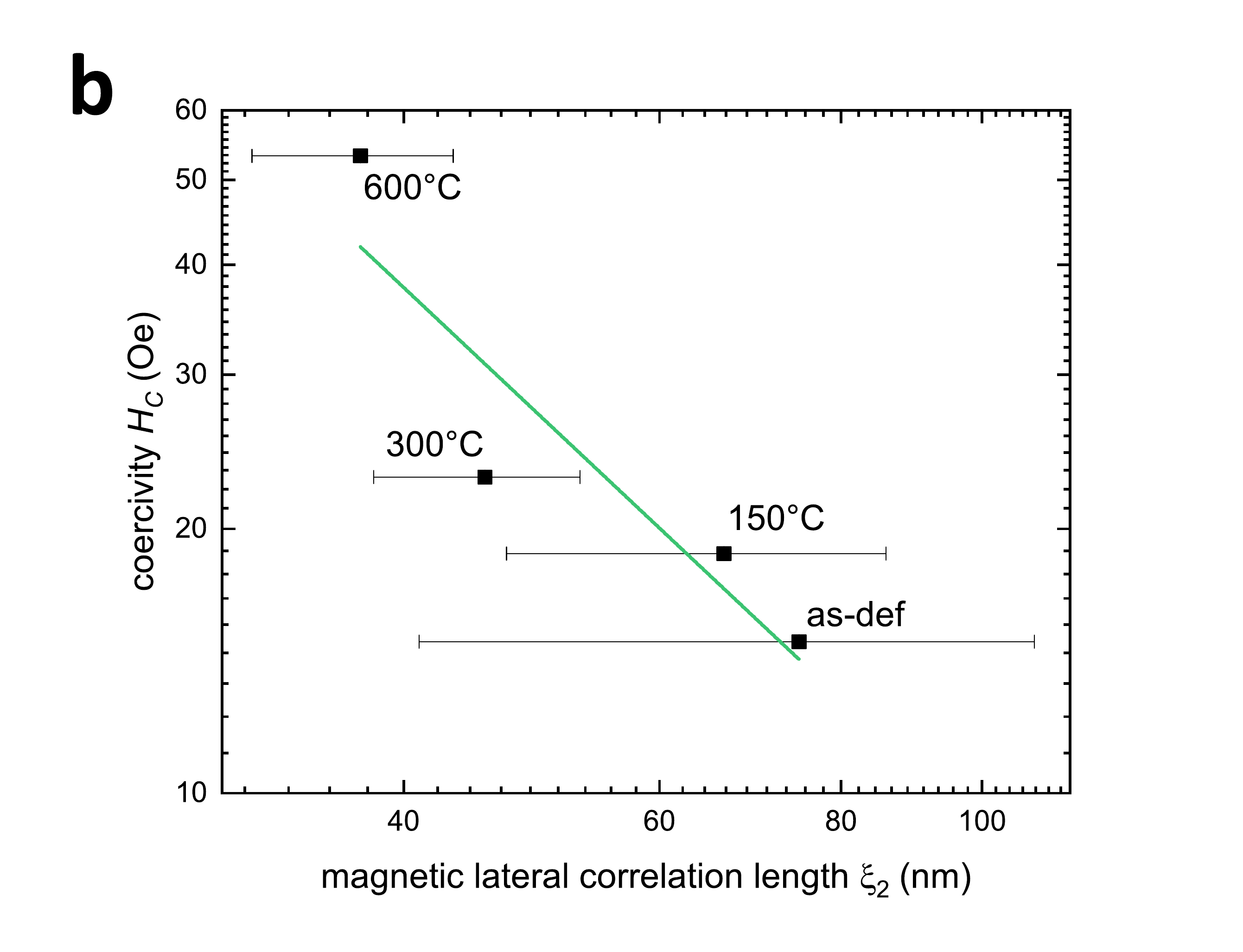}
\caption{(a) Magnetic lateral correlation lengths $\xi_1$, $\xi_2$ in comparison to topographic $\xi_{TOPO}$ as a function of annealing temperature (c.f. tab.~S.1). The data are slightly shifted in x for better visibility. (b) Coercivity $H_C$ determined by SQUID-magnetometry versus magnetic $\xi_2$ obtained from correlating analysis of the MFM images. Note the log-axes.}
\label{fig:xi}
\end{figure}
\section{Conclusion}
In this study, Co-Cu samples with high Co-fractions are prepared using SPD by HPT. The thermal stability of Co72-Cu28 (wt.\%) is investigated in detail, with the main focus being on microstructural and magnetic properties.
By using a two-step HPT deformation at different temperatures a homogenization of strain distribution, and as a consequence an improved chemical homogeneity, is obtained. The resulting sample is nanocrystalline and exhibits a supersaturated, fcc single-phase microstructure. The microstructure exhibits a remarkeable thermal stability, since no changes can be observed in SEM and XRD for annealing treatments up to 400\degree C. For annealing at 600\degree C, a demixed state, consisting of two fcc-phases, is present. Magnetometric measurements showed an increasing coercivity as a function of annealing temperature, meaning the observed changes in coercivity cannot be explained by formation of multidomain structures. Correlating MFM measurements with DC-SQUID magnetometry reveals domain sizes close to single-domain, whereas the demixed phases (with a magnetic correlation length smaller than the grain size) yield a magnetic hardening effect, which is able to explain the observed increase in coercivity -- an indicator of demixing of the phases in context with spinodal decomposition.
\section*{Acknowledgments}
This project has received funding from the European Research Council (ERC) under the European Union’s Horizon 2020 research and innovation programme (Grant No. 757333). A. M. acknowledges the support from the Lise Meitner fellowship by Austrian Science Fund (FWF): M 2323-N36.
\bibliography{libraryjsamd}

\begin{thebibliography}{10}

\bibitem{GLEITER1989223}
H~Gleiter.
\newblock Nanocrystalline materials.
\newblock {\em Prog. Mater. Sci.}, 33(4):223 -- 315, 1989.

\bibitem{herzer2005round}
G~Herzer, M~Vazquez, M~Knobel, A~Zhukov, T~Reininger, HA~Davies, Roland
  Gr{\"o}ssinger, and JL~Sanchez Ll.
\newblock Round table discussion: Present and future applications of
  nanocrystalline magnetic materials.
\newblock {\em J. Magn. Magn. Mater.}, 294(2):252--266, 2005.

\bibitem{suzuki1991soft}
K~Suzuki, A~Makino, A~Inoue, and T~Masumoto.
\newblock Soft magnetic properties of nanocrystalline bcc {Fe-Zr-B} and
  {Fe-M-B-Cu} ({M}= transition metal) alloys with high saturation
  magnetization.
\newblock {\em J. Appl. Phys.}, 70(10):6232--6237, 1991.

\bibitem{herzer2013modern}
G~Herzer.
\newblock Modern soft magnets: Amorphous and nanocrystalline materials.
\newblock {\em Acta Mater.}, 61(3):718--734, 2013.

\bibitem{ambrose1993magnetic}
T~Ambrose, A~Gavrin, and CL~Chien.
\newblock Magnetic properties of metastable fcc {Fe-Cu} alloys prepared by high
  energy ball milling.
\newblock {\em J. Magn. Magn. Mater.}, 124(1-2):15--19, 1993.

\bibitem{shen2005soft}
T~D Shen, R~B Schwarz, and J~D Thompson.
\newblock Soft magnetism in mechanically alloyed nanocrystalline materials.
\newblock {\em Phys. Rev. B}, 72(1):014431, 2005.

\bibitem{valiev2000bulk}
R~Z Valiev, R~K Islamgaliev, and I~V Alexandrov.
\newblock Bulk nanostructured materials from severe plastic deformation.
\newblock {\em Prog. Mater. Sci.}, 45(2):103--189, 2000.

\bibitem{pippan2006limits}
R~Pippan, F~Wetscher, M~Hafok, A~Vorhauer, and I~Sabirov.
\newblock The limits of refinement by severe plastic deformation.
\newblock {\em Adv. Eng. Mater.}, 8(11):1046--1056, 2006.

\bibitem{KILMAMETOV2018337}
A~R Kilmametov, Y~Ivanisenko, A~A Mazilkin, B~B Straumal, A~S Gornakova, O~B
  Fabrichnaya, M~J Kriegel, D~Rafaja, and H~Hahn.
\newblock The $\alpha\rightarrow\omega$ and $\beta\rightarrow\omega$ phase
  transformations in {Ti–Fe} alloys under high-pressure torsion.
\newblock {\em Acta Mater.}, 144:337 -- 351, 2018.

\bibitem{o1989opportunities}
R~C {O'Handley}.
\newblock Opportunities in magnetic anisotropy and magnetostriction.
\newblock {\em Mater. Sci. Eng. B}, 3(4):365--369, 1989.

\bibitem{kormout2017deformation}
K~S Kormout, R~Pippan, and A~Bachmaier.
\newblock Deformation-induced supersaturation in immiscible material systems
  during high-pressure torsion.
\newblock {\em Adv. Eng. Mater.}, 19(4):1600675, 2017.

\bibitem{stuckler2019magnetic}
M~St{\"u}ckler, H~Krenn, R~Pippan, L~Weissitsch, S~Wurster, and A~Bachmaier.
\newblock Magnetic binary supersaturated solid solutions processed by severe
  plastic deformation.
\newblock {\em Nanomaterials}, 9(1):6, 2019.

\bibitem{BACHMAIER2017744}
A~Bachmaier, H~Krenn, P~Knoll, H~Aboulfadl, and R~Pippan.
\newblock Tailoring the magnetic properties of nanocrystalline cu-co alloys
  prepared by high-pressure torsion and isothermal annealing.
\newblock {\em J. Alloys Compd}, 725:744 -- 749, 2017.

\bibitem{bachmaier2016process}
A~Bachmaier, J~Schmauch, H~Aboulfadl, A~Verch, and C~Motz.
\newblock On the process of co-deformation and phase dissolution in a hard-soft
  immiscible {CuCo} alloy system during high-pressure torsion deformation.
\newblock {\em Acta Mater.}, 115:333--346, 2016.

\bibitem{pippan2010saturation}
R~Pippan, S~Scheriau, A~Taylor, M~Hafok, A~Hohenwarter, and A~Bachmaier.
\newblock Saturation of fragmentation during severe plastic deformation.
\newblock {\em Annu. Rev. Mater. Sci.}, 40:319--343, 2010.

\bibitem{renk2019saturation}
O~Renk and R~Pippan.
\newblock Saturation of grain refinement during severe plastic deformation of
  single phase materials: reconsiderations, current status and open questions.
\newblock {\em Mater. Trans.}, 60(7):1270--1282, 2019.

\bibitem{renk2015increasing}
O~Renk, A~Hohenwarter, K~Eder, K~S Kormout, J~M Cairney, and R~Pippan.
\newblock Increasing the strength of nanocrystalline steels by annealing: Is
  segregation necessary?
\newblock {\em Scr. Mater.}, 95:27--30, 2015.

\bibitem{nelson1945experimental}
J~B Nelson and D~P Riley.
\newblock An experimental investigation of extrapolation methods in the
  derivation of accurate unit-cell dimensions of crystals.
\newblock {\em Proc. Phys. Soc.}, 57(3):160, 1945.

\bibitem{vegard1921konstitution}
L~Vegard.
\newblock {Die Konstitution der Mischkristalle und die Raumf{\"u}llung der
  Atome}.
\newblock {\em Z. Phys.}, 5(1):17--26, 1921.

\bibitem{kriegel2019thermal}
M~J Kriegel, A~Kilmametov, V~Klemm, C~Schimpf, B~B Straumal, A~S Gornakova,
  Y~Ivanisenko, O~Fabrichnaya, H~Hahn, and D~Rafaja.
\newblock Thermal stability of athermal $\omega$-ti (fe) produced upon
  quenching of $\beta$-ti (fe).
\newblock {\em Adv. Eng. Mater.}, 21(1):1800158, 2019.

\bibitem{childress1991reentrant}
J~R Childress and C~L Chien.
\newblock Reentrant magnetic behavior in fcc {Co-Cu} alloys.
\newblock {\em Phys. Rev. B}, 43(10):8089, 1991.

\bibitem{qdusa1500-011}
Quantum Design, San Diego, California.
\newblock {\em Application Note 1500-011}, 5 2010.
\newblock Rev. A0.

\bibitem{kronmuller2003micromagnetism}
H~Kronm{\"u}ller and M~F{\"a}hnle.
\newblock {\em Micromagnetism and the microstructure of ferromagnetic solids}.
\newblock Cambridge university press, 2003.

\bibitem{jamet2001magnetic}
M~Jamet, W~Wernsdorfer, C~Thirion, D~Mailly, V~Dupuis, P~M{\'e}linon, and
  A~P{\'e}rez.
\newblock Magnetic anisotropy of a single cobalt nanocluster.
\newblock {\em Phys. Rev. Lett.}, 86(20):4676, 2001.

\bibitem{andreeva2016study}
N~V Andreeva, A~V Filimonov, A~I Rudskoi, G~S Burkhanov, I~S Tereshina, G~A
  Politova, and I~A Pelevin.
\newblock A study of nanostructure magnetosolid {Nd-Ho-Fe-Co-B} materials via
  atomic force microscopy and magnetic force microscopy.
\newblock {\em Phys. Solid State}, 58(9):1862--1869, 2016.

\bibitem{zhao2000characterization}
Y~Zhao, G-C Wang, and T-M Lu.
\newblock {\em Characterization of Amorphous and Crystalline Rough
  Surface--Principles and Applications}.
\newblock Elsevier, 2000.

\bibitem{teichert2002self}
C~Teichert.
\newblock Self-organization of nanostructures in semiconductor heteroepitaxy.
\newblock {\em Phys. Rep.}, 365(5-6):335--432, 2002.

\bibitem{teichert2009ion}
C~Teichert, J~J De~Miguel, and T~Bobek.
\newblock Ion beam sputtered nanostructured semiconductor surfaces as templates
  for nanomagnet arrays.
\newblock {\em J. Phys. Condens. Matter}, 21(22):224025, 2009.

\bibitem{tolan1998evidence}
M~Tolan, O~H Seeck, J-P Schlomka, W~Press, J~Wang, S~K Sinha, Z~Li, M~H
  Rafailovich, and J~Sokolov.
\newblock Evidence for capillary waves on dewetted polymer film surfaces: A
  combined {X}-ray and atomic force microscopy study.
\newblock {\em Phys. Rev. Lett.}, 81(13):2731, 1998.

\bibitem{bachmaier2015phase}
A~Bachmaier, M~Pfaff, M~Stolpe, H~Aboulfadl, and C~Motz.
\newblock Phase separation of a supersaturated nanocrystalline {Cu-Co} alloy
  and its influence on thermal stability.
\newblock {\em Acta Mater.}, 96:269--283, 2015.

\bibitem{wurster2020GMR}
S~Wurster, M~Stückler, L~Weissitsch, T~Müller, and A~Bachmaier.
\newblock Microstructural changes influencing the magnetoresistive behavior of
  bulk nanocrystalline materials.
\newblock {\em Appl. Sci.}, 10(15):5094, Jul 2020.

\end{thebibliography}
\newpage
\section*{Supplementary}
\begin{table}[h!]
\renewcommand\thetable{S.1} 
   \centering
\caption{Roughness $\sigma$, lateral correlation length $\xi$, and Hurst parameter $\alpha$ as evaluated from auto-correlation function $C(x)$ and height-height correlation function $H(x)$ for the topographic (top) and magnetic signal (bottom). The data represent the mean values from at least 6 individual scans. As the magnetic signal exhibits two different Hurst parameters, the magnetic $C(x)$ is fitted by [eq.~7] with $c$ being a proportionality constant.}
\begin{tabular}{llScSScSScSScS}
\hline
\multicolumn{2}{|c|}{fit-parameters} & \multicolumn{3}{c|}{as-deformed} & \multicolumn{3}{c|}{150\degree C} & \multicolumn{3}{c|}{300\degree C} & \multicolumn{3}{c|}{600\degree C}                \\ 
\hline
\multicolumn{1}{|l}{topographic} & \multicolumn{1}{l|}{$\sigma_{TOPO}$ (nm)}      & 3.2  & $\pm$                & \multicolumn{1}{S|}{0.7}  & 3.0    & $\pm$ & \multicolumn{1}{S|}{0.2}  & 2.6 & $\pm$ & \multicolumn{1}{S|}{0.5}  & 2.8  & $\pm$ & \multicolumn{1}{S|}{0.2}  \\
\multicolumn{1}{|l}{}           & \multicolumn{1}{l|}{$\xi_{TOPO}$ (nm)}         & 75   & $\pm$                & \multicolumn{1}{S|}{30}   & 55   & $\pm$ & \multicolumn{1}{S|}{10}    & 55  & $\pm$ & \multicolumn{1}{S|}{15}   & 70   & $\pm$ & \multicolumn{1}{S|}{20}   \\
\multicolumn{1}{|l}{}           & \multicolumn{1}{l|}{$\alpha_{TOPO}$}           & 0.55 & $\pm$                & \multicolumn{1}{S|}{0.05} & 0.35 & $\pm$ & \multicolumn{1}{S|}{0.05} & 0.60 & $\pm$ & \multicolumn{1}{S|}{0.05} & 0.65 & $\pm$ & \multicolumn{1}{S|}{0.10} \\
 \hline
\multicolumn{1}{|l}{magnetic}   & \multicolumn{1}{l|}{$\sigma$ (deg)} & 0.15  & $\pm$                & \multicolumn{1}{S|}{0.03}   & 0.12  & $\pm$ & \multicolumn{1}{S|}{0.02}   & 0.11 & $\pm$ & \multicolumn{1}{S|}{0.01}   & 0.17  & $\pm$ & \multicolumn{1}{S|}{0.01}   \\
\multicolumn{1}{|l}{}           & \multicolumn{1}{l|}{$\xi_1$ (nm)}       & 110  & $\pm$                & \multicolumn{1}{S|}{50}   & 110  & $\pm$ & \multicolumn{1}{S|}{70}   & 20  & $\pm$ & \multicolumn{1}{S|}{15}   & 20   & $\pm$ & \multicolumn{1}{S|}{10}   \\
\multicolumn{1}{|l}{}           & \multicolumn{1}{l|}{$\xi_2$ (nm)}       & 80   & $\pm$ & \multicolumn{1}{S|}{30}   & 70   & $\pm$ & \multicolumn{1}{S|}{20}   & 50  & $\pm$ & \multicolumn{1}{S|}{10}    & 40   & $\pm$ & \multicolumn{1}{S|}{10}    \\
\multicolumn{1}{|l}{}           & \multicolumn{1}{l|}{$\alpha_1$}         & 0.40  & $\pm$                & \multicolumn{1}{S|}{0.10}  & 0.30  & $\pm$ & \multicolumn{1}{S|}{0.10}  & 0.15 & $\pm$ & \multicolumn{1}{S|}{0.05}  & 0.25  & $\pm$ & \multicolumn{1}{S|}{0.05}  \\
\multicolumn{1}{|l}{}           & \multicolumn{1}{l|}{$\alpha_2$}         & 0.25  & $\pm$                & \multicolumn{1}{S|}{0.10}  & 0.30  & $\pm$ & \multicolumn{1}{S|}{0.10}  & 0.35 & $\pm$ & \multicolumn{1}{S|}{0.10}  & 0.60  & $\pm$ & \multicolumn{1}{S|}{0.05}  \\
\multicolumn{1}{|l}{}           & \multicolumn{1}{l|}{$c$}               & 0.2  & $\pm$                & \multicolumn{1}{S|}{0.1}  & 0.2  & $\pm$ & \multicolumn{1}{S|}{0.2}  & 0.1 & $\pm$ & \multicolumn{1}{S|}{0.2}  & 0.2  & $\pm$ & \multicolumn{1}{S|}{0.1}  \\ \hline
\end{tabular}
\end{table}
\subsection*{M-H curves}
Figure.~S.2 shows the hysteresis of the as-deformed and the annealed states in comparison to the pure Co-powder, measured at 8~K. The as-deformed and the annealed samples are measured in hard axis configuration and exhibit therefore a smaller susceptibility than the powder, which morphological shape appears more uniform. The evolution of coercivity and saturation magnetization have already been discussed in detail in fig.~5.
\begin{figure}[h!]
\includegraphics[width=0.7\linewidth]{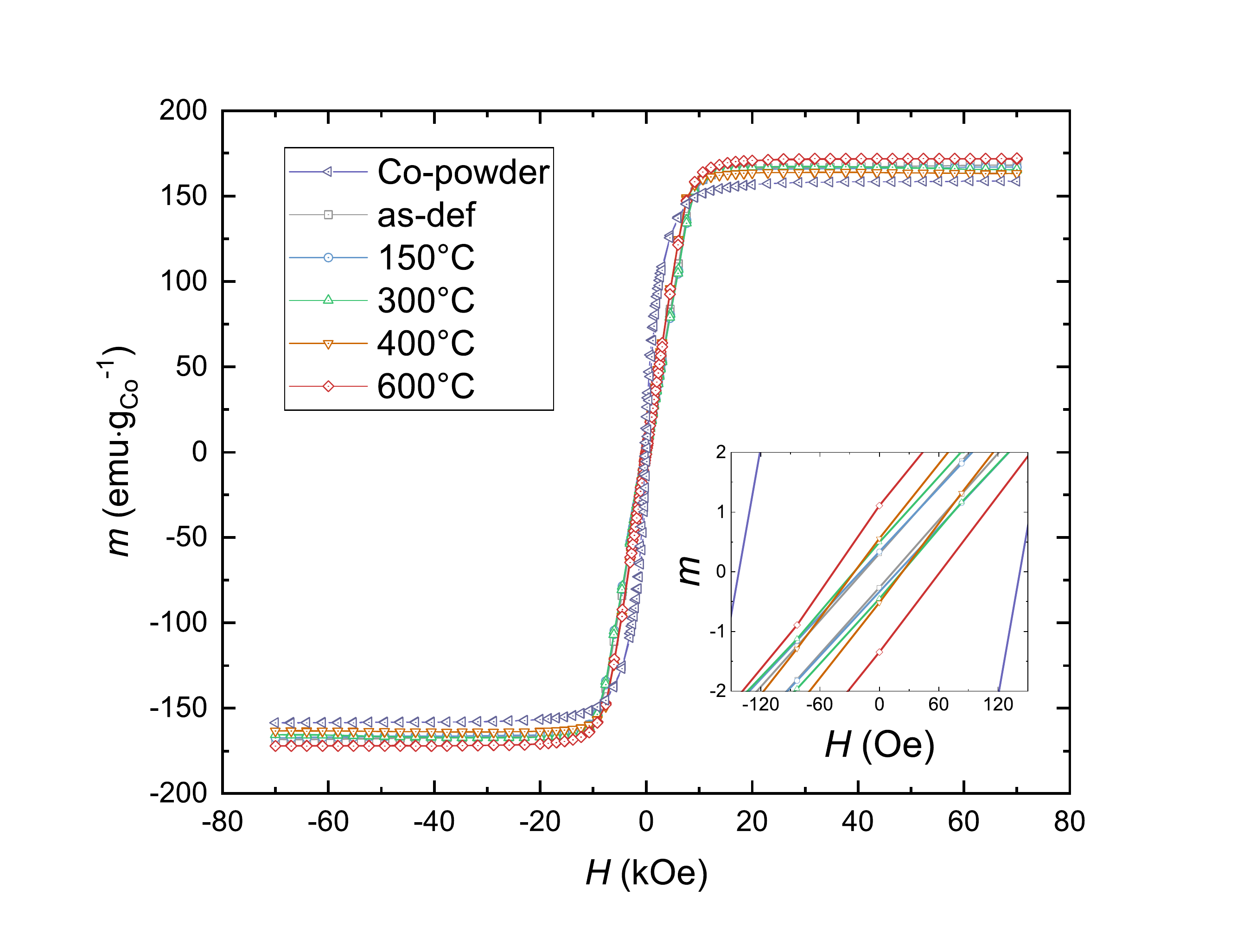}
\renewcommand\thefigure{S.2} 
   \centering
\caption{Hysteresis loops measured by SQUID-magnetometry at 8~K. The inset shows parts of the hysteresis loops at small fields.}
\end{figure}
\end{document}